\documentclass[aps,prd,twocolumn,amsmath,amssymb,amsfonts,nofootinbib,superscriptaddress]{revtex4-1}
\usepackage[utf8]{inputenc}
\usepackage{graphicx}% Include figure files
\usepackage{subfigure}
\usepackage{datatool}
\usepackage[scientific-notation=true]{siunitx}
\usepackage[dvipsnames]{xcolor}
\usepackage{lineno}

\DTLsetseparator{,}

\DTLloaddb{sensdb}{parameters/sens_fig_parms.txt}
\DTLassign{sensdb}{1}{\tobs=tobs}
\DTLassign{sensdb}{1}{\fstep=fstep}
\DTLassign{sensdb}{1}{\fmin=fmin}
\DTLassign{sensdb}{1}{\fmax=fmax}
\DTLassign{sensdb}{1}{\fstep=fstep}
\DTLassign{sensdb}{1}{\mcone=mc1}
\DTLassign{sensdb}{1}{\mctwo=mc2}
\DTLassign{sensdb}{1}{\mcthree=mc3}
\DTLassign{sensdb}{1}{\mcthree=mc3}
\DTLassign{sensdb}{1}{\gaaama=Gamma}
\DTLassign{sensdb}{1}{\thetathr=theta_thr}
\DTLassign{sensdb}{1}{\crthr=CR_thr}

\DTLloaddb{origsour}{parameters/orig_sour.txt}
\DTLassign{origsour}{1}{\adjfnaught=adj_f0}
\DTLassign{origsour}{1}{\adjxnaught=adj_x0}
\DTLassign{origsour}{1}{\origmc=mc}
\DTLassign{origsour}{1}{\orighnaught=h0}
\DTLassign{origsour}{1}{\origtnaught=t00}
\DTLassign{origsour}{1}{\origk=kn}
\DTLassign{origsour}{1}{\adjtnaught=adj_t0}
\DTLassign{origsour}{1}{\dur=dur}
\DTLassign{origsour}{1}{\detnameinj=det}

\DTLloaddb{hmjob}{parameters/hm_job.txt}

\DTLloaddb{candsour}{parameters/cand_sour.txt}
\DTLassign{candsour}{1}{\candfnaught=f0}
\DTLassign{candsour}{1}{\candmc=mc}
\DTLassign{candsour}{1}{\candk=kn}
\DTLassign{candsour}{1}{\binsaway=distaway}
\DTLassign{candsour}{1}{\tfftused=tfft}
\DTLassign{candsour}{1}{\candtnaught=fepoch}

\newcommand{\bea}{\begin{eqnarray}}
\newcommand{\eea}{\end{eqnarray}}
\newcommand{\be}{\begin{equation}}
\newcommand{\ee}{\end{equation}}

\newcommand{\TFFT}{T_\text{FFT}}

\newcommand{\Tobs}{T_\text{obs}}

\newcommand{\round}{
\num[round-precision=6,round-mode=figures,
     scientific-notation=false]}
 
\newcommand{\roundtwo}{
\num[round-precision=2,round-mode=figures,
     scientific-notation=true]}

\newcommand{\roundfour}{
\num[round-precision=4,round-mode=figures,
     scientific-notation=false]}

\def\erfc{\mathrm{erfc}}

\begin{document}
\begin{filecontents*}{parameters/sens_fig_parms.txt}
tobs,mc1,mc2,mc3,fmin,fmax,fstep,Gamma,theta_thr,CR_thr
31536000,0.001,0.0001,1e-05,50,2000,50,0.9,2.5,5
\end{filecontents*}

\begin{filecontents*}{parameters/orig_sour.txt}
a,d,v_a,v_d,pepoch,f0,df0,ddf0,fepoch,t00,eps,eta,psi,cosiota,h0,snr,coord,chphase,dfrsim,x0,kn,tau,n,del_n,dur,lam,beta,mc,adj_x0,adj_f0,adj_t0,det
328.815308210047,23.8246643205737,0,0,58581.8453472222,140,0.000428908864966966,0,58581.8453472222,58581.8453472222,1,0.976905977211516,-39.8703206039313,-0.804919190001181,1e-22,1,0,0,-0.2,1.89230834615565e-06,-5.79734160655099e-12,122403.62531104,3.66666666666667,-0,20520,340.709973937565,33.9883193921549,0.00100664050709327,1.73424962459646e-06,144.654899054071,58581.9636805556,Hanford
\end{filecontents*}

\begin{filecontents*}{parameters/hm_job.txt}
minf,maxf,df,dur,patch_1,patch_2,n,ref_perc_time,frenh,gridk_1,gridk_2,gridk_3,gridk_4,gridk_5,gridk_6,gridk_7,gridk_8,gridk_9,gridk_10,gridk_11,gridk_12,gridk_13,gridk_14,gridk_15,gridk_16,gridk_17,gridk_18,gridk_19,gridk_20,gridk_21,gridk_22,gridk_23,gridk_24,gridk_25,gridk_26,gridk_27,gridk_28,gridk_29,gridk_30,gridk_31,gridk_32,gridk_33,gridk_34,gridk_35,gridk_36,gridk_37,gridk_38,gridk_39,gridk_40,gridk_41,gridk_42,gridk_43,gridk_44,gridk_45,gridk_46,gridk_47,gridk_48,gridk_49,gridk_50,gridk_51,gridk_52,gridk_53,gridk_54,gridk_55,gridk_56,gridk_57,gridk_58,gridk_59,gridk_60,gridk_61,gridk_62,gridk_63,gridk_64,gridk_65,gridk_66,gridk_67,gridk_68,gridk_69,gridk_70,gridk_71,gridk_72,gridk_73,gridk_74,gridk_75,gridk_76,gridk_77,gridk_78,gridk_79,gridk_80,gridk_81,gridk_82,gridk_83,gridk_84,gridk_85,gridk_86,gridk_87,gridk_88,gridk_89,gridk_90,gridk_91,gridk_92,gridk_93,gridk_94,gridk_95,gridk_96,gridk_97,gridk_98,gridk_99,gridk_100,gridk_101,gridk_102,gridk_103,gridk_104,gridk_105,gridk_106,gridk_107,gridk_108,gridk_109,gridk_110,gridk_111,gridk_112,gridk_113,gridk_114,gridk_115,gridk_116,gridk_117,gridk_118,gridk_119,gridk_120,gridk_121,gridk_122,gridk_123,gridk_124,gridk_125,gridk_126,gridk_127,gridk_128,gridk_129,gridk_130,gridk_131,gridk_132,gridk_133,gridk_134,gridk_135,gridk_136,gridk_137,gridk_138,gridk_139,gridk_140,gridk_141,gridk_142,gridk_143,gridk_144,gridk_145,gridk_146,gridk_147,gridk_148,gridk_149,gridk_150,gridk_151,gridk_152,gridk_153,gridk_154,gridk_155,gridk_156,gridk_157,gridk_158,gridk_159,gridk_160,gridk_161,gridk_162,gridk_163,gridk_164,gridk_165,gridk_166,gridk_167,gridk_168,gridk_169,gridk_170,gridk_171,gridk_172,gridk_173,gridk_174,gridk_175,gridk_176,gridk_177,gridk_178,gridk_179,gridk_180,gridk_181,gridk_182,gridk_183,gridk_184,gridk_185,gridk_186,gridk_187,gridk_188,gridk_189,gridk_190,gridk_191,gridk_192,gridk_193,gridk_194,gridk_195,gridk_196,gridk_197,gridk_198,gridk_199,gridk_200,gridk_201,gridk_202,gridk_203,gridk_204,gridk_205,gridk_206,gridk_207,gridk_208,gridk_209,gridk_210,gridk_211,gridk_212,gridk_213,gridk_214,gridk_215,gridk_216,gridk_217,gridk_218,gridk_219,gridk_220,gridk_221,gridk_222,gridk_223,gridk_224,gridk_225,gridk_226,gridk_227,gridk_228,gridk_229,gridk_230,gridk_231,gridk_232,gridk_233,gridk_234,gridk_235,gridk_236,gridk_237,gridk_238,gridk_239,gridk_240,gridk_241,gridk_242,gridk_243,gridk_244,gridk_245,gridk_246,gridk_247,gridk_248,gridk_249,gridk_250,gridk_251,gridk_252,gridk_253,gridk_254,gridk_255,gridk_256,gridk_257,gridk_258,gridk_259,gridk_260,gridk_261,gridk_262,gridk_263,gridk_264,gridk_265,gridk_266,gridk_267,gridk_268,gridk_269,gridk_270,gridk_271,gridk_272,gridk_273,gridk_274,gridk_275,gridk_276,gridk_277,gridk_278,gridk_279,gridk_280,gridk_281,gridk_282,gridk_283,gridk_284,gridk_285,gridk_286,gridk_287,gridk_288,gridk_289,gridk_290,gridk_291,gridk_292,gridk_293,gridk_294,gridk_295,gridk_296,gridk_297,gridk_298,gridk_299,gridk_300,gridk_301,gridk_302,gridk_303,gridk_304,gridk_305,gridk_306,gridk_307,gridk_308,gridk_309,gridk_310,gridk_311,gridk_312,gridk_313,gridk_314,gridk_315,gridk_316,gridk_317,gridk_318,gridk_319,gridk_320,gridk_321,gridk_322,gridk_323,gridk_324,gridk_325,gridk_326,gridk_327,gridk_328,gridk_329,gridk_330,gridk_331,gridk_332,gridk_333,gridk_334,gridk_335,gridk_336,gridk_337,gridk_338,gridk_339,gridk_340,gridk_341,gridk_342,gridk_343,gridk_344,gridk_345,gridk_346,gridk_347,gridk_348,gridk_349,gridk_350,gridk_351,gridk_352,gridk_353,gridk_354,gridk_355,gridk_356,gridk_357,gridk_358,gridk_359,gridk_360,gridk_361,gridk_362,gridk_363,gridk_364,gridk_365,gridk_366,gridk_367,gridk_368,gridk_369,gridk_370,gridk_371,gridk_372,gridk_373,gridk_374,gridk_375,gridk_376,gridk_377,gridk_378,gridk_379,gridk_380,gridk_381,gridk_382,gridk_383,gridk_384,gridk_385,gridk_386,gridk_387,gridk_388,gridk_389,gridk_390,gridk_391,gridk_392,gridk_393,gridk_394,gridk_395,gridk_396,gridk_397,gridk_398,gridk_399,gridk_400,gridk_401,gridk_402,gridk_403,gridk_404,gridk_405,gridk_406,gridk_407,gridk_408,gridk_409,gridk_410,gridk_411,gridk_412,gridk_413,gridk_414,gridk_415,gridk_416,gridk_417,gridk_418,gridk_419,gridk_420,gridk_421,gridk_422,gridk_423,gridk_424,gridk_425,gridk_426,gridk_427,gridk_428,gridk_429,gridk_430,gridk_431,gridk_432,gridk_433,gridk_434,gridk_435,gridk_436,gridk_437,gridk_438,gridk_439,gridk_440,gridk_441,gridk_442,gridk_443,gridk_444,gridk_445,gridk_446,gridk_447,gridk_448,gridk_449,gridk_450,gridk_451,gridk_452,gridk_453,gridk_454,gridk_455,gridk_456,gridk_457,gridk_458,gridk_459,gridk_460,gridk_461,gridk_462,gridk_463,gridk_464,gridk_465,gridk_466,gridk_467,gridk_468,gridk_469,gridk_470,gridk_471,gridk_472,gridk_473,gridk_474,gridk_475,gridk_476,gridk_477,gridk_478,gridk_479,gridk_480,gridk_481,gridk_482,gridk_483,gridk_484,gridk_485,gridk_486,gridk_487,gridk_488,gridk_489,gridk_490,gridk_491,gridk_492,gridk_493,gridk_494,gridk_495,gridk_496,gridk_497,gridk_498,gridk_499,gridk_500,gridk_501,gridk_502,gridk_503,gridk_504,gridk_505,gridk_506,gridk_507,gridk_508,gridk_509,gridk_510,gridk_511,gridk_512,gridk_513,gridk_514,gridk_515,gridk_516,gridk_517,gridk_518,gridk_519,gridk_520,gridk_521,gridk_522,gridk_523,gridk_524,gridk_525,gridk_526,gridk_527,gridk_528,gridk_529,gridk_530,gridk_531,gridk_532,gridk_533,gridk_534,gridk_535,gridk_536,gridk_537,gridk_538,gridk_539,gridk_540,gridk_541,gridk_542,gridk_543,gridk_544,gridk_545,gridk_546,gridk_547,gridk_548,gridk_549,gridk_550,gridk_551,gridk_552,gridk_553,gridk_554,gridk_555,gridk_556,gridk_557,gridk_558,gridk_559,gridk_560,gridk_561,gridk_562,gridk_563,gridk_564,gridk_565,gridk_566,gridk_567,gridk_568,gridk_569,gridk_570,gridk_571,gridk_572,gridk_573,gridk_574,gridk_575,gridk_576,gridk_577,gridk_578,gridk_579,gridk_580,gridk_581,gridk_582,gridk_583,gridk_584,gridk_585,gridk_586,gridk_587,gridk_588,gridk_589,gridk_590,gridk_591,gridk_592,gridk_593,gridk_594,gridk_595,gridk_596,gridk_597,gridk_598,gridk_599,gridk_600,gridk_601,gridk_602,gridk_603,gridk_604,gridk_605,gridk_606,gridk_607,gridk_608,gridk_609,gridk_610,gridk_611,gridk_612,gridk_613,gridk_614,gridk_615,gridk_616,gridk_617,gridk_618,gridk_619,gridk_620,gridk_621,gridk_622,gridk_623,gridk_624,gridk_625,gridk_626,gridk_627,gridk_628,gridk_629,gridk_630,gridk_631,gridk_632,gridk_633,gridk_634,gridk_635,gridk_636,gridk_637,gridk_638,gridk_639,gridk_640,gridk_641,gridk_642,gridk_643,gridk_644,gridk_645,gridk_646,gridk_647,gridk_648,gridk_649,gridk_650,gridk_651,gridk_652,gridk_653,gridk_654,gridk_655,gridk_656,gridk_657,gridk_658,gridk_659,gridk_660,gridk_661,gridk_662,gridk_663,gridk_664,epoch,dx
140,150,0.03125,20520,340.709973937565,33.9883193921549,3.66666666666667,0.5,1,4.4995065999909e-12,4.50294467795504e-12,4.50638538295835e-12,4.50982871700815e-12,4.51327468211331e-12,4.51672328028421e-12,4.52017451353278e-12,4.52362838387249e-12,4.52708489331835e-12,4.53054404388688e-12,4.5340058375962e-12,4.5374702764659e-12,4.54093736251718e-12,4.54440709777274e-12,4.54787948425684e-12,4.55135452399529e-12,4.55483221901544e-12,4.55831257134621e-12,4.56179558301803e-12,4.56528125606291e-12,4.56876959251443e-12,4.57226059440768e-12,4.57575426377933e-12,4.57925060266761e-12,4.5827496131123e-12,4.58625129715473e-12,4.58975565683781e-12,4.59326269420599e-12,4.5967724113053e-12,4.60028481018333e-12,4.60379989288922e-12,4.60731766147368e-12,4.61083811798901e-12,4.61436126448906e-12,4.61788710302923e-12,4.62141563566653e-12,4.62494686445951e-12,4.62848079146832e-12,4.63201741875466e-12,4.63555674838182e-12,4.63909878241465e-12,4.6426435229196e-12,4.64619097196469e-12,4.64974113161951e-12,4.65329400395525e-12,4.65684959104466e-12,4.6604078949621e-12,4.66396891778349e-12,4.66753266158636e-12,4.6710991284498e-12,4.67466832045451e-12,4.67824023968278e-12,4.68181488821847e-12,4.68539226814706e-12,4.68897238155561e-12,4.69255523053276e-12,4.69614081716878e-12,4.69972914355551e-12,4.70332021178639e-12,4.70691402395648e-12,4.71051058216242e-12,4.71410988850245e-12,4.71771194507644e-12,4.72131675398584e-12,4.7249243173337e-12,4.7285346372247e-12,4.73214771576513e-12,4.73576355506285e-12,4.73938215722738e-12,4.74300352436982e-12,4.7466276586029e-12,4.75025456204094e-12,4.75388423679991e-12,4.75751668499738e-12,4.76115190875252e-12,4.76478991018615e-12,4.7684306914207e-12,4.77207425458021e-12,4.77572060179036e-12,4.77936973517843e-12,4.78302165687336e-12,4.7866763690057e-12,4.79033387370761e-12,4.7939941731129e-12,4.79765726935702e-12,4.80132316457702e-12,4.80499186091161e-12,4.80866336050112e-12,4.81233766548752e-12,4.81601477801443e-12,4.81969470022707e-12,4.82337743427235e-12,4.82706298229878e-12,4.83075134645653e-12,4.8344425288974e-12,4.83813653177486e-12,4.841833357244e-12,4.84553300746156e-12,4.84923548458594e-12,4.85294079077718e-12,4.85664892819698e-12,4.86035989900867e-12,4.86407370537727e-12,4.86779034946941e-12,4.87150983345341e-12,4.87523215949923e-12,4.87895732977849e-12,4.88268534646448e-12,4.88641621173214e-12,4.89014992775807e-12,4.89388649672054e-12,4.89762592079949e-12,4.9013682021765e-12,4.90511334303485e-12,4.90886134555947e-12,4.91261221193695e-12,4.91636594435558e-12,4.9201225450053e-12,4.92388201607773e-12,4.92764435976615e-12,4.93140957826554e-12,4.93517767377255e-12,4.9389486484855e-12,4.94272250460439e-12,4.94649924433092e-12,4.95027886986844e-12,4.95406138342201e-12,4.95784678719836e-12,4.96163508340592e-12,4.9654262742548e-12,4.9692203619568e-12,4.9730173487254e-12,4.97681723677579e-12,4.98062002832483e-12,4.9844257255911e-12,4.98823433079485e-12,4.99204584615805e-12,4.99586027390434e-12,4.99967761625909e-12,5.00349787544936e-12,5.00732105370389e-12,5.01114715325315e-12,5.0149761763293e-12,5.01880812516621e-12,5.02264300199947e-12,5.02648080906634e-12,5.03032154860584e-12,5.03416522285867e-12,5.03801183406724e-12,5.04186138447568e-12,5.04571387632984e-12,5.04956931187728e-12,5.05342769336729e-12,5.05728902305085e-12,5.0611533031807e-12,5.06502053601125e-12,5.06889072379869e-12,5.07276386880089e-12,5.07663997327747e-12,5.08051903948976e-12,5.08440106970082e-12,5.08828606617546e-12,5.0921740311802e-12,5.0960649669833e-12,5.09995887585474e-12,5.10385576006625e-12,5.10775562189129e-12,5.11165846360506e-12,5.1155642874845e-12,5.11947309580828e-12,5.12338489085682e-12,5.12729967491227e-12,5.13121745025855e-12,5.1351382191813e-12,5.13906198396791e-12,5.14298874690753e-12,5.14691851029105e-12,5.15085127641112e-12,5.15478704756212e-12,5.15872582604021e-12,5.16266761414328e-12,5.166612414171e-12,5.17056022842478e-12,5.17451105920779e-12,5.17846490882497e-12,5.18242177958301e-12,5.18638167379036e-12,5.19034459375726e-12,5.19431054179567e-12,5.19827952021937e-12,5.20225153134387e-12,5.20622657748645e-12,5.21020466096618e-12,5.2141857841039e-12,5.2181699492222e-12,5.22215715864547e-12,5.22614741469987e-12,5.23014071971333e-12,5.23413707601556e-12,5.23813648593805e-12,5.24213895181409e-12,5.24614447597872e-12,5.25015306076879e-12,5.25416470852293e-12,5.25817942158155e-12,5.26219720228686e-12,5.26621805298284e-12,5.27024197601528e-12,5.27426897373175e-12,5.27829904848163e-12,5.28233220261607e-12,5.28636843848805e-12,5.29040775845232e-12,5.29445016486543e-12,5.29849566008575e-12,5.30254424647344e-12,5.30659592639046e-12,5.31065070220057e-12,5.31470857626937e-12,5.31876955096423e-12,5.32283362865433e-12,5.32690081171069e-12,5.33097110250612e-12,5.33504450341523e-12,5.33912101681449e-12,5.34320064508214e-12,5.34728339059825e-12,5.35136925574472e-12,5.35545824290527e-12,5.35955035446542e-12,5.36364559281253e-12,5.3677439603358e-12,5.37184545942621e-12,5.37595009247661e-12,5.38005786188166e-12,5.38416877003785e-12,5.3882828193435e-12,5.39240001219878e-12,5.39652035100566e-12,5.40064383816798e-12,5.40477047609139e-12,5.4089002671834e-12,5.41303321385334e-12,5.41716931851241e-12,5.42130858357361e-12,5.42545101145182e-12,5.42959660456374e-12,5.43374536532795e-12,5.43789729616483e-12,5.44205239949666e-12,5.44621067774753e-12,5.45037213334341e-12,5.45453676871211e-12,5.45870458628329e-12,5.46287558848849e-12,5.46704977776108e-12,5.4712271565363e-12,5.47540772725126e-12,5.47959149234492e-12,5.48377845425811e-12,5.48796861543353e-12,5.49216197831573e-12,5.49635854535114e-12,5.50055831898806e-12,5.50476130167665e-12,5.50896749586896e-12,5.5131769040189e-12,5.51738952858226e-12,5.5216053720167e-12,5.52582443678176e-12,5.53004672533888e-12,5.53427224015134e-12,5.53850098368435e-12,5.54273295840495e-12,5.54696816678212e-12,5.55120661128669e-12,5.55544829439139e-12,5.55969321857083e-12,5.56394138630153e-12,5.56819280006188e-12,5.57244746233219e-12,5.57670537559464e-12,5.58096654233331e-12,5.5852309650342e-12,5.58949864618518e-12,5.59376958827605e-12,5.59804379379849e-12,5.60232126524609e-12,5.60660200511435e-12,5.61088601590067e-12,5.61517330010436e-12,5.61946386022666e-12,5.62375769877068e-12,5.62805481824148e-12,5.63235522114601e-12,5.63665890999315e-12,5.6409658872937e-12,5.64527615556036e-12,5.64958971730776e-12,5.65390657505246e-12,5.65822673131294e-12,5.66255018860958e-12,5.66687694946472e-12,5.67120701640262e-12,5.67554039194944e-12,5.6798770786333e-12,5.68421707898426e-12,5.68856039553427e-12,5.69290703081726e-12,5.69725698736907e-12,5.70161026772749e-12,5.70596687443224e-12,5.71032681002499e-12,5.71469007704934e-12,5.71905667805084e-12,5.72342661557699e-12,5.72779989217724e-12,5.73217651040296e-12,5.73655647280751e-12,5.74093978194617e-12,5.74532644037619e-12,5.74971645065676e-12,5.75410981534903e-12,5.75850653701612e-12,5.76290661822309e-12,5.76731006153698e-12,5.77171686952676e-12,5.7761270447634e-12,5.78054058981981e-12,5.78495750727087e-12,5.78937779969344e-12,5.79380146966633e-12,5.79822851977034e-12,5.80265895258822e-12,5.80709277070471e-12,5.81152997670652e-12,5.81597057318234e-12,5.82041456272283e-12,5.82486194792065e-12,5.8293127313704e-12,5.83376691566871e-12,5.83822450341416e-12,5.84268549720733e-12,5.84714989965078e-12,5.85161771334908e-12,5.85608894090876e-12,5.86056358493835e-12,5.8650416480484e-12,5.86952313285141e-12,5.87400804196191e-12,5.87849637799642e-12,5.88298814357345e-12,5.88748334131352e-12,5.89198197383915e-12,5.89648404377487e-12,5.90098955374719e-12,5.90549850638467e-12,5.91001090431783e-12,5.91452675017924e-12,5.91904604660346e-12,5.92356879622708e-12,5.92809500168868e-12,5.93262466562887e-12,5.93715779069028e-12,5.94169437951756e-12,5.94623443475738e-12,5.95077795905841e-12,5.95532495507139e-12,5.95987542544905e-12,5.96442937284615e-12,5.96898679991948e-12,5.97354770932788e-12,5.97811210373219e-12,5.98267998579531e-12,5.98725135818217e-12,5.99182622355971e-12,5.99640458459695e-12,6.00098644396491e-12,6.00557180433667e-12,6.01016066838736e-12,6.01475303879415e-12,6.01934891823623e-12,6.02394830939488e-12,6.02855121495338e-12,6.03315763759712e-12,6.03776758001348e-12,6.04238104489194e-12,6.046998034924e-12,6.05161855280325e-12,6.05624260122532e-12,6.0608701828879e-12,6.06550130049073e-12,6.07013595673565e-12,6.07477415432652e-12,6.07941589596929e-12,6.08406118437198e-12,6.08871002224468e-12,6.09336241229952e-12,6.09801835725075e-12,6.10267785981466e-12,6.10734092270962e-12,6.1120075486561e-12,6.11667774037662e-12,6.12135150059579e-12,6.1260288320403e-12,6.13070973743895e-12,6.13539421952258e-12,6.14008228102414e-12,6.14477392467868e-12,6.14946915322331e-12,6.15416796939727e-12,6.15887037594185e-12,6.16357637560047e-12,6.16828597111863e-12,6.17299916524393e-12,6.17771596072606e-12,6.18243636031684e-12,6.18716036677016e-12,6.19188798284202e-12,6.19661921129056e-12,6.20135405487597e-12,6.2060925163606e-12,6.21083459850889e-12,6.21558030408738e-12,6.22032963586475e-12,6.22508259661178e-12,6.22983918910137e-12,6.23459941610853e-12,6.23936328041041e-12,6.24413078478626e-12,6.24890193201748e-12,6.25367672488756e-12,6.25845516618215e-12,6.26323725868902e-12,6.26802300519805e-12,6.27281240850128e-12,6.27760547139286e-12,6.2824021966691e-12,6.28720258712841e-12,6.29200664557138e-12,6.29681437480072e-12,6.30162577762127e-12,6.30644085684003e-12,6.31125961526614e-12,6.31608205571089e-12,6.32090818098771e-12,6.32573799391219e-12,6.33057149730206e-12,6.33540869397721e-12,6.34024958675968e-12,6.34509417847367e-12,6.34994247194555e-12,6.35479447000382e-12,6.35965017547918e-12,6.36450959120445e-12,6.36937272001464e-12,6.37423956474693e-12,6.37911012824066e-12,6.38398441333733e-12,6.38886242288063e-12,6.39374415971642e-12,6.39862962669271e-12,6.40351882665972e-12,6.40841176246982e-12,6.41330843697758e-12,6.41820885303975e-12,6.42311301351524e-12,6.42802092126516e-12,6.43293257915282e-12,6.4378479900437e-12,6.44276715680548e-12,6.44769008230801e-12,6.45261676942335e-12,6.45754722102577e-12,6.46248143999169e-12,6.46741942919979e-12,6.47236119153089e-12,6.47730672986805e-12,6.48225604709651e-12,6.48720914610374e-12,6.4921660297794e-12,6.49712670101534e-12,6.50209116270566e-12,6.50705941774664e-12,6.51203146903678e-12,6.51700731947681e-12,6.52198697196966e-12,6.52697042942048e-12,6.53195769473664e-12,6.53694877082774e-12,6.5419436606056e-12,6.54694236698425e-12,6.55194489287998e-12,6.55695124121126e-12,6.56196141489884e-12,6.56697541686566e-12,6.57199325003693e-12,6.57701491734007e-12,6.58204042170474e-12,6.58706976606285e-12,6.59210295334854e-12,6.59713998649819e-12,6.60218086845044e-12,6.60722560214616e-12,6.61227419052847e-12,6.61732663654274e-12,6.62238294313659e-12,6.6274431132599e-12,6.63250714986479e-12,6.63757505590564e-12,6.64264683433909e-12,6.64772248812406e-12,6.65280202022168e-12,6.6578854335954e-12,6.66297273121089e-12,6.6680639160361e-12,6.67315899104127e-12,6.67825795919888e-12,6.68336082348369e-12,6.68846758687274e-12,6.69357825234533e-12,6.69869282288306e-12,6.70381130146979e-12,6.70893369109166e-12,6.7140599947371e-12,6.71919021539682e-12,6.72432435606381e-12,6.72946241973336e-12,6.73460440940304e-12,6.7397503280727e-12,6.74490017874451e-12,6.7500539644229e-12,6.75521168811462e-12,6.76037335282871e-12,6.76553896157652e-12,6.77070851737167e-12,6.77588202323012e-12,6.78105948217012e-12,6.7862408972122e-12,6.79142627137925e-12,6.79661560769643e-12,6.80180890919122e-12,6.80700617889342e-12,6.81220741983514e-12,6.81741263505081e-12,6.82262182757718e-12,6.82783500045331e-12,6.8330521567206e-12,6.83827329942275e-12,6.84349843160581e-12,6.84872755631815e-12,6.85396067661045e-12,6.85919779553575e-12,6.86443891614941e-12,6.86968404150912e-12,6.87493317467491e-12,6.88018631870916e-12,6.88544347667656e-12,6.89070465164418e-12,6.8959698466814e-12,6.90123906485996e-12,6.90651230925396e-12,6.91178958293982e-12,6.91707088899634e-12,6.92235623050465e-12,6.92764561054825e-12,6.93293903221298e-12,6.93823649858705e-12,6.94353801276104e-12,6.94884357782786e-12,6.95415319688281e-12,6.95946687302354e-12,6.96478460935009e-12,6.97010640896484e-12,6.97543227497255e-12,6.98076221048036e-12,6.98609621859779e-12,6.99143430243671e-12,6.99677646511138e-12,7.00212270973846e-12,7.00747303943697e-12,7.01282745732831e-12,7.01818596653627e-12,7.02354857018704e-12,7.02891527140919e-12,7.03428607333367e-12,7.03966097909383e-12,7.04503999182542e-12,7.05042311466659e-12,7.05581035075786e-12,7.06120170324218e-12,7.0665971752649e-12,7.07199676997374e-12,7.07740049051887e-12,7.08280834005284e-12,7.08822032173061e-12,7.09363643870956e-12,7.09905669414948e-12,7.10448109121256e-12,7.10990963306344e-12,7.11534232286914e-12,7.12077916379913e-12,7.12622015902527e-12,7.13166531172188e-12,7.13711462506568e-12,7.14256810223583e-12,7.14802574641391e-12,7.15348756078393e-12,7.15895354853235e-12,7.16442371284805e-12,7.16989805692234e-12,7.17537658394899e-12,7.18085929712419e-12,7.18634619964659e-12,7.19183729471726e-12,7.19733258553975e-12,7.20283207532002e-12,7.20833576726652e-12,7.2138436645901e-12,7.21935577050412e-12,7.22487208822436e-12,7.23039262096906e-12,7.23591737195893e-12,7.24144634441714e-12,7.24697954156931e-12,7.25251696664354e-12,7.25805862287038e-12,7.26360451348287e-12,7.2691546417165e-12,7.27470901080925e-12,7.28026762400156e-12,7.28583048453635e-12,7.29139759565902e-12,7.29696896061745e-12,7.30254458266201e-12,7.30812446504552e-12,7.31370861102333e-12,7.31929702385325e-12,7.32488970679559e-12,7.33048666311314e-12,7.3360878960712e-12,7.34169340893755e-12,7.34730320498247e-12,7.35291728747875e-12,7.35853565970167e-12,7.36415832492901e-12,7.36978528644106e-12,7.37541654752062e-12,7.38105211145299e-12,7.38669198152599e-12,7.39233616102995e-12,7.3979846532577e-12,7.4036374615046e-12,7.40929458906852e-12,7.41495603924987e-12,7.42062181535156e-12,7.42629192067902e-12,7.43196635854023e-12,7.43764513224568e-12,7.44332824510839e-12,7.44901570044392e-12,7.45470750157035e-12,7.46040365180832e-12,7.46610415448097e-12,58581.9636805556,8.74616106323137e-10
\end{filecontents*}

\begin{filecontents*}{parameters/cand_sour.txt}
lam,bet,f0,df0,ddf0,n,x0,kn,dur,tau,fepoch,pepoch,v_a,v_d,a,d,ulam,ubet,ux,uk,tfft,det,cr,mc,distaway
340.709973937565,33.9883193921549,144.646795924941,-0.000483959803404397,0,3.66666666666667,1.73450871096114e-06,-5.80265895258822e-12,20520,112080.689533068,58581.9636805556,58581.9636805556,0,0,328.815308210047,23.8246643205737,0,0,8.74616106323137e-10,4.43381811649092e-15,32,Hanford,22.0586481481481,0.00100719438245189,1.23531392352236
\end{filecontents*}

% \title{Continuous gravitational waves as probes of planetary-mass PBHs}
\title{Probing planetary-mass primordial black holes with continuous gravitational waves}
% \title{Detecting transient continuous gravitational waves from %{galactic} 
% inspiraling {planetary-mass} PBHs}%\\using continuous gravitational wave methods}
% \author{UCLouvain Virgo}
\author{Andrew L. Miller}
\email{andrew.miller@uclouvain.be}
\affiliation{Université catholique de Louvain, B-1348 Louvain-la-Neuve, Belgium}
\author{Sébastien Clesse}
\email{sebastien.clesse@ulb.ac.be}
\affiliation{Service de Physique Th\'eorique, Universit\'e Libre de Bruxelles, Boulevard du Triomphe CP225, B-1050 Brussels, Belgium}
\author{Federico De Lillo}
\affiliation{Université catholique de Louvain, B-1348 Louvain-la-Neuve, Belgium}
\author{Giacomo Bruno}
\affiliation{Université catholique de Louvain, B-1348 Louvain-la-Neuve, Belgium}
\author{Antoine Depasse}
\affiliation{Université catholique de Louvain, B-1348 Louvain-la-Neuve, Belgium}
\author{Andres Tanasijczuk}
\affiliation{Université catholique de Louvain, B-1348 Louvain-la-Neuve, Belgium}
\date{\today}

\begin{abstract}
Gravitational waves can probe the existence of planetary-mass primordial black holes. Considering a mass range of $[10^{-7}-10^{-2}]M_\odot$, inspiraling primordial black holes could emit either continuous gravitational waves, quasi-monochromatic signals that last for many years, or transient continuous waves, signals whose frequency evolution follows a power law and last for $\mathcal{O}$(hours-months).
We show that primordial black hole binaries in our galaxy may 
produce detectable gravitational waves for different mass functions and formation mechanisms. In order to detect these inspirals, we adapt methods originally designed to search for gravitational waves from asymmetrically rotating neutron stars. The first method, the Frequency-Hough, exploits the continuous, quasi-monochromatic nature of inspiraling black holes that are sufficiently light and far apart such that their orbital frequencies can be approximated as linear with a small spin-up. The second method, the Generalized Frequency-Hough, drops the assumption of linearity and allows the signal frequency to follow a power-law evolution. We explore the parameter space to which each method is sensitive, derive a theoretical sensitivity estimate, determine optimal search parameters and calculate the computational cost of all-sky and directed searches.  We forecast limits on the abundance of primordial black holes within our galaxy, showing that we can constrain the fraction of dark matter that primordial black holes compose, $f_{\rm PBH}$, to be $f_{\rm PBH}\lesssim 1$ for chirp masses between $[4\times 10^{-5}-10^{-3}]M_\odot$ for current detectors. For the Einstein Telescope, we expect the constraints to improve to $f_{\rm PBH}\lesssim 10^{-2}$ for chirp masses between [$10^{-4}-10^{-3}]M_\odot$.

\end{abstract}

\maketitle

\section{Introduction}
The rates, progenitor masses and low effective spins of the black hole mergers detected by LIGO/Virgo~\cite{aasi2015advanced,acernese2014advanced,Abbott:2016blz,TheLIGOScientific:2016pea,Abbott:2016nmj,Abbott:2017vtc,Abbott:2017oio,Abbott:2017gyy,LIGOScientific:2018mvr,LIGOScientific:2020stg,Abbott:2020uma,Abbott:2020khf,Abbott:2020tfl,Abbott:2020mjq} have renewed interest in primordial black holes (PBHs) in the $ [1-100] M_\odot$ range~\cite{Bird:2016dcv,Clesse:2016vqa,Sasaki:2016jop}.  There exists a variety of PBH formation scenarios, e.g. the gravitational collapse of inhomogeneities that can be generated during inflation, reheating or phase transitions, that allow PBHs to compose a fraction ~\cite{Sasaki:2016jop,Ali-Haimoud:2017rtz,Hall:2020daa,DeLuca:2020jug} or all of dark matter in the Universe~\cite{Bird:2016dcv,Clesse:2016vqa,clesse2018seven,Carr:2019kxo,Jedamzik:2020ypm,Jedamzik:2020omx,Boehm:2020jwd,DeLuca:2020agl}. In addition to being responsible for forming stellar-mass black holes, the collapse process may also have triggered Baryogenesis in the Universe in pockets surrounding each PBH, and linked the observed baryon-to-photon ratio to the abundance of PBHs at formation~\cite{Garcia-Bellido:2019vlf,Carr:2019hud}. We refer the reader to~\cite{Carr:2016drx,Carr:2009jm} for reviews on PBHs and~\cite{Carr:2020gox,Carr:2020xqk,Green:2020jor} for a summary of recent developments.  

Given the increasing interest for PBHs and novel stellar models to explain the unexpected properties of black hole mergers (e.g. low spins, the masses of GW190814 and GW190521), it is now crucial to find ways to experimentally distinguish primordial and astrophysical black holes.  In this context, detecting sub-solar black holes would almost certainly point to a primordial origin\footnote{See however~\cite{kouvaris2018nonprimordial,Dasgupta:2020mqg} for another sub-solar black hole formation channel, in a particular dark matter scenario, with specific spin predictions.}. However, different theories of PBH formation predict a vast range of masses that span several orders of magnitude, e.g. if curvature fluctuations at the origin of PBH formation are nearly scale invariant~\cite{Carr:2019kxo,Byrnes:2018clq,Jedamzik:2020ypm,Jedamzik:2020omx,DeLuca:2020agl} -- a natural prediction of inflation -- or come from a broad peak in the PBH power spectrum~\cite{Clesse:2015wea,Ezquiaga:2017fvi}. The wide range of possible PBH masses underscores the need to develop a variety of methods to probe the existence of PBHs.

If the aforementioned theories are correct, the known thermal history of the Universe would have left imprints in the PBH mass function~\cite{Niemeyer:1997mt,Jedamzik:1996mr,Byrnes:2018clq,Carr:2019kxo}, independently of the mechanism responsible for the pre-existing curvature fluctuations.  In particular, the equation of state of the Universe varies at the Quantum Chromodynamics transition at $\sim 100$ MeV, and at higher energies $\sim 100$ GeV when the standard model top quark, W and Z bosons and Higgs boson became non-relativistic.  This transition induces transient variations of the overdensity threshold leading to gravitational collapse, easing PBH formation and resulting in two unique features in the mass function: (1) a high peak at the solar mass scale and (2) two bumps at $\sim 30 M_\odot$ and $\sim 10^{-5} M_\odot $.  Such a mass function could explain a series of puzzling observations~\cite{clesse2018seven,Carr:2019kxo}, such as unexpected microlensing events, LIGO/Virgo black hole mergers, spatial correlations in the source-subtracted infrared and soft X-ray backgrounds, some properties of dwarf galaxies, and the existence of super-massive black holes at high redshifts. NANOGrav's recent observation of a possible stochastic gravitational-wave background at nano-Hertz frequencies~\cite{Arzoumanian:2020vkk} may also hint a the existence of stellar~\cite{DeLuca:2020agl,Vaskonen:2020lbd} or planetary-mass PBHs~\cite{Domenech:2020ers}.
Moreover, the planetary-mass range is specifically motivated by recent detections of star and quasar microlensing events~\cite{Niikura:2019kqi,Hawkins:2020zie,bhatiani2019confirmation,Hawkins:2020rqu,mroz2017no}, suggesting that PBHs or compact objects with masses between $10^{-6}M_\odot$ and $ 10^{-5}M_\odot$ make up a fraction of the dark matter $f_{\rm PBH} \sim \mathcal{O}(0.01)$.  This fraction is more than expected for free-floating planets, but is consistent with PBHs in the unified scenario presented in~\cite{Carr:2019kxo}.  It has even been recently suggested that the hypothetical Planet 9 could be a PBH of mass $\sim 10^{-6} M_\odot$~\cite{Scholtz:2019csj} captured by the solar system, and detection strategies based on the accretion of small Oort Cloud objects have been proposed~\cite{Siraj:2020upy}.  
However all these observations and their derived limits are subject to large astrophysical uncertainties, for instance due to the clustering properties of PBHs~\cite{Garcia-Bellido:2017xvr,Calcino:2018mwh,Belotsky:2018wph,Carr:2019kxo,Carr:2019kxo,Trashorras:2020mwn,DeLuca:2020jug}. It is therefore important to find a complementary way to probe the existence of such objects, and to distinguish PBHs from other sources.  

In this paper, we show that LIGO/Virgo and the third generation gravitational-wave detectors, e.g. Einstein Telescope~\cite{Punturo:2010zza,Hild:2010id,Maggiore:2019uih}, can detect gravitational waves from nearby galactic PBH binaries.  Because their expected merging rate is several orders of magnitudes larger than stellar-mass binaries, we show that this novel method could set new limits on the abundance of PBHs in the planetary-mass range.  
At such small masses, the inspiral phase of the PBH mergers could last for potentially thousands years, or hours-months, depending on the chirp mass and orbital frequency. Therefore, from a data analysis point of view, the signal's phase evolution is actually closer to that from asymmetrically rotating neutron stars than from canonical binary system signals. From most neutron stars, we expect continuous gravitational waves, quasi-monochromatic, quasi-infinite signals, arising from a small deformation on the star's surface \cite{Osborne:2019iph,ushomirsky2000deformations,PhysRevD.88.044004,abbott2019all} or from accretion from a companion star \cite{watts2008detecting}. From newborn neutron stars, we expect transient continuous gravitational waves, shorter signals, $\mathcal{O}$(hours-days), whose frequency evolution follows a power law \cite{lasky2017braking,Sarin:2018vsi}. Both signal types are much simpler than those searched for in traditional matched filter and burst searches for compact binaries coalescences, e.g. \cite{usman2016pycbc}. Indeed, the inspiral orbital frequency can be modelled as a power law until the inner-most stable circular orbit, which, for these small masses, occurs at frequencies far outside of the detector sensitivity band (see section \ref{gwinsp} and \cite{maggiore2008gravitational}). Moreover, when the chirp mass is small enough, the power-law frequency evolution reduces to a linear frequency behavior in time. The simplicity of the signal phase evolution means that transient continuous-wave and continuous-wave techniques \cite{riles2017recent,sieniawska2019continuous} can be applied to search for inspiraling planetary-mass PBHs. 
% However, we will show that continuous-wave techniques alone are not sufficient to cover the entire accessible parameter space, due primarily to the requirement of linear frequency evolution over time \cite{Astone:2014esa,Krishnan:2004sv}, and therefore that transient continuous wave methods are needed to cover the rest of it, but with a high computational cost. 

We can envision performing continuous-wave-like searches for inspiraling PBHs. Since we do not know the masses of the primordial black a priori, nor do we have particular electromagnetic observations to guide us, \emph{targeted} searches \cite{abbott2019searches}, in which we correct exactly for the phase evolution of the signal by knowing its sky location, frequency and spin-up, are not possible
% { (SC: if fact, they are possible, in the direction of objects dense in dark matter, such as some ultra-faint dwarf galaxies)}. 
But a \emph{directed} search \cite{piccinni2020directed,aasi2015searches,longpmr,Keitel:2019zhb} for PBHs pointing towards a known location, e.g. the galactic center, or where a high concentration of PBHs could exist, is possible, in which we would search over a range of frequencies and spin-ups of these systems. Additionally, an \emph{all-sky}  search \cite{abbott2019all}, where we assume no knowledge of the signal, is also plausible, since PBHs could be inspiraling anywhere in the sky. Many continuous-wave methods have already been adapted to search for gravitational waves from dark matter candidates, e.g. the axion, around black holes \cite{isi2019directed,palomba2019direct,d2018semicoherent}, and dark matter particles that interact directly with the interferometers \cite{guo2019searching,pierce2019dark,Miller:2020vsl}. For planetary-mass PBH insprials, we show that directed searches are more computationally feasible than all-sky ones, and are still astrophysically interesting.

% continuous-wave searches are typically divided in three categories depending upon how much knowledge we have about the sources we are looking for. \emph{Targeted} searches are performed when the frequency, spin-down and sky position of are known \cite{abbott2019searches}; \emph{directed} searches are done when the sky position is known but no other orbital parameters are \cite{piccinni2020directed,aasi2015searches}; \emph{all-sky} searches assume no knowledge about any source and analyze each sky direction separately looking for isolated or binary neutron stars \cite{abbott2019all}. Recently, transient continuous wave searches have also been performed looking for gravitational waves from a remnant of GW170817 \cite{longpmr} and after pulsar glitches \cite{keitel2019first}. Furthermore, many continuous-wave methods have already been adapted to search for gravitational waves from dark matter candidates, e.g. the axion, around black holes \cite{isi2019directed,palomba2019direct,d2018semicoherent} and dark matter particles that interact directly with the interferometers \cite{guo2019searching,pierce2019dark,Miller:2020vsl}. We explore the merits of performing each type of search for PBH binaries in this paper as well.

% Though continuous-wave techniques can be applied to detect PBH mergers, it is important to evaluate how likely such mergers are in the Milky Way.  The rates expected for the two main binary formation channels are analyzed below.

The paper is structured as follows: in section \ref{gwinsp} we describe the expected signal and the accessible parameter space. We then explain in section \ref{pbhmerge} the main formation channels for PBH mergers and evaluate how likely they are to happen in the Milky Way. We report the methods to search for these signals in section \ref{searchmethod}. In section \ref{anamethod}, we perform a sophisticated analysis of the methods, estimate the computation time of directed and all-sky searches and derive a theoretical estimate of the methods' sensitivity. Finally we make some concluding remarks in section \ref{concl}.

% The paper is structured as follows: in section \ref{pbhmerge} we describe the main formation channels for PBH mergers and evaluate how likely they are to happen in the Milky Way.  We describe the expected signal and the accessible parameter space in section \ref{gwinsp} and the methods to search for these signals in section \ref{searchmethod}. In section \ref{anamethod}, we perform a sophisticated analysis of the methods, estimate the computation time of directed and all-sky searches and derive a A theoretical estimate of the methods' sensitivity. Finally we make some concluding remarks in section \ref{concl}.

\section{Gravitational Waves from inspirals: The Signal}\label{gwinsp}

% This section focuses on describing the gravitational waves emitted by PBH mergers. In contrast to standard binary black hole and neutron star systems, the signal is easy to model, so long as the PBH system maintains a quasi-circular orbit. We describe the expected signal and the situations under which the assumption of quasi-circular orbits is valid in section \ref{sig}, and then we explain briefly the re

% \subsection{The signal}\label{sig}
The inspiral of two black holes, many orbits away from the innermost stable circular orbit, can be approximated as two point masses in a circular orbit around their center of mass (see section 4.1 of \cite{maggiore2008gravitational}), whose orbital frequency $\omega_{\rm orb}$ is given by Kepler's law. When accounting for the loss of orbital energy due to gravitational-wave emission, the distance between the two black holes decreases, which means that   $\omega_{\rm orb}$ increases. Equating the power lost due to gravitational-wave emission with the rate of change of the orbital energy of the system, and knowing that the gravitational-wave frequency $f_{\rm gw}=\pi \omega_{\rm orb}$ 
% {(SC: is this correct?  shouldn't it be divided by $4\pi$?)} %%% fgw=2*f_orb ; \omega_orb=2*pi*f_orb --> fgw=pi*forb
we arrive at \cite{maggiore2008gravitational}:

\begin{equation}
    \dot{f}_{\rm gw}=\frac{96}{5}\pi^{8/3}\left(\frac{G\mathcal{M}}{c^3}\right)^{5/3} f_{\rm gw}^{11/3},
    \label{chirp_pl}
\end{equation}
where $\dot{f}_{\rm gw}$ is the rate of change of the frequency (the spin-up), $\mathcal{M}$ is the chirp mass of the system, $c$ is the speed of light, and $G$ is Newton's gravitational constant.  For the most massive PBHs we consider, the gravitational-wave frequency at the innermost stable circular orbit is $O$(MHz), ensuring that approximation of the inspiral as a circular orbit, and therefore Equation~\ref{chirp_pl} is valid for the full range of LIGO/Virgo frequencies. 

Equation \ref{chirp_pl} is a power law, with a braking index $n=11/3$ and a constant of proportionality $k$:
% {
\be
{-k}\equiv\frac{96}{5}\pi^{8/3}\left(\frac{G\mathcal{M}}{c^3}\right)^{5/3}~,
\ee
% } 
where we have written $-k$ to be consistent with our notation in section \ref{ght}.
This type of signal can be searched for with techniques developed to detect transient continuous waves lasting $\mathcal{O}$(hours-days) that come from remnants of binary neutron star mergers or supernova \cite{PhysRevD.98.102004,Oliver:2018dpt}. Transient continuous waves also follow power laws, but $n=5$ or $n=7$ for canonical gravitational-wave emission from a deformation \cite{Sarin:2018vsi} or r-modes \cite{mytidis2015constraining,mytidis2015sensitivity,owen1998gravitational}, respectively, and $\dot{f}$ is negative. 

Integrating Equation \ref{chirp_pl}, we obtain the frequency evolution:
\begin{equation}
f_{\rm gw}(t)=f_0\left[1-\frac{8}{3}kf_0^{8/3}(t-t_0)\right]^{-\frac{3}{8}}~,
\label{powlaws}
\end{equation}
where $t_0$ is a reference time for the gravitational-wave frequency $f_0$ and $t-t_0=t_{\rm merg}$ is the time to merger. We also solve equation \ref{powlaws} for $t_{\rm merg}$: 

\begin{equation}
    % t_{\rm merg}=\frac{15}{768}\pi^{-8/3}\left(\frac{c^3}{G\mathcal{M}}\right)^{-5/3} (f-f_0)^{-8/3 }
    t_{\rm merg}=\frac{3}{8}\frac{f^{-8/3}-f_0^{-8/3 }}{k}.
\end{equation}
With some modifications to the existing pipeline in \cite{PhysRevD.98.102004}, we can search for signals that follow equation \ref{powlaws}. 
% I also got your comment about the targeted pbh search- 'targeted' has a specific meaning in CW- you know everything, f , dot{f}, phase evolution - is it true in case you mention  No, you do not know f because there might be many PBH binary sources in these environemnts...  So we can let targeted... ok ok I will explain more clearly to avoid confusion.  Great!  

We must also consider the amplitude evolution of the signal over time, which differs from the continuous-wave case \cite{maggiore2008gravitational}:

\begin{equation}
h(t)=\frac{4}{d}\left(\frac{G \mathcal{M}}{c^2}\right)^{5/3}\left(\frac{\pi f_{\rm gw}(t)}{c}\right)^{2/3},
\label{hoft}
\end{equation}
where $d$ is the distance to the source.

\begin{figure}
    \centering
    \hspace{-3mm} %\includegraphics[width=0.45\textwidth]{figures/time_before_merg.png}
    \includegraphics[width=0.49\textwidth]{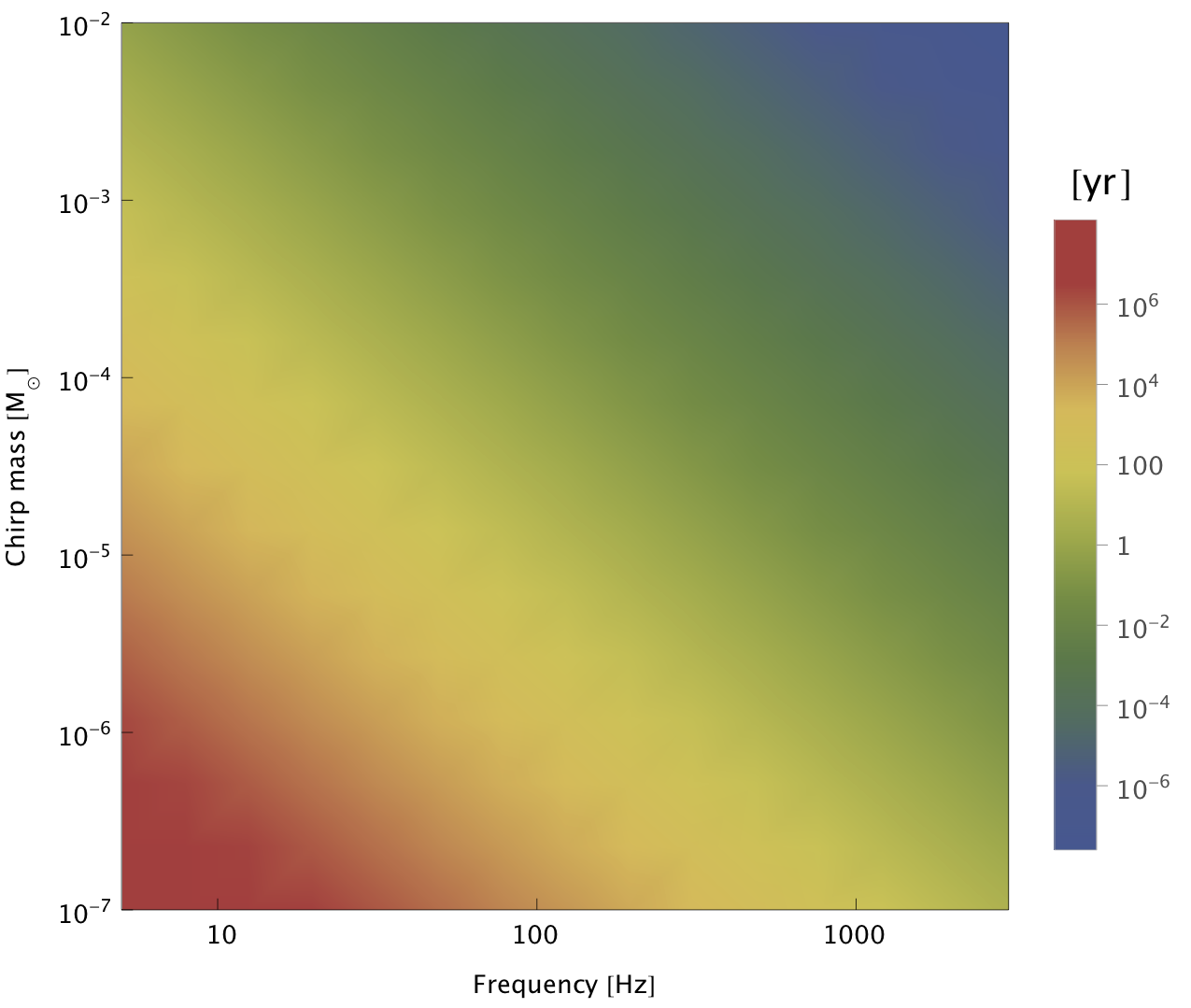}  
%    \caption{Time before merging as a function of the frequency, for different chirp masses.}
    \caption{Time before merger as a function of frequency and chirp mass. The widely distributed signal durations in the parameter space imply that different techniques are needed to probe the existence of PBHs at different masses.}
    \label{fig:tmerg}
\end{figure}

Due to their extremely small masses, PBHs are expected to inspiral for very long times compared to typical LIGO/Virgo binary black hole signals, potentially for months or years. $t_{\rm merg}$ is shown in Figure~\ref{fig:tmerg} as a function of the frequency and chirp mass, which demonstrates that PBH inspirals are ideal candidates for continuous-wave and transient continuous-wave searches. However, depending on the PBH masses, the spin-ups could fall outside of the range typically analyzed in continuous-wave searches \cite{abbott2019all}. Moreover if Equation \ref{chirp_pl} cannot be approximated to be linear (see section \ref{linvspow}), continuous-wave searches are blind to PBH mergers even if their spin-ups are small. 

\begin{figure}
    \centering
    \includegraphics[width=\columnwidth]{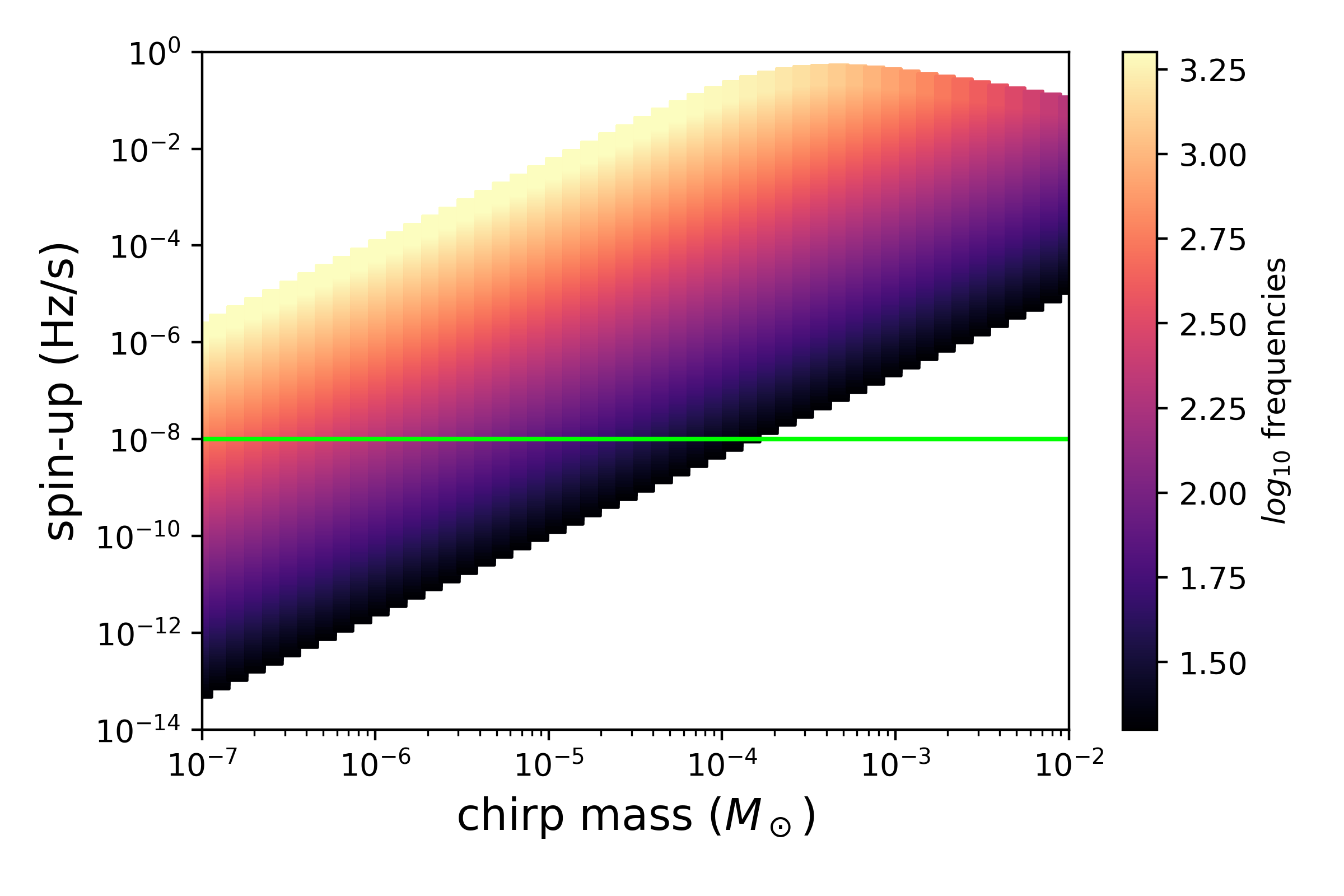}
    \caption{Spin-up as a function of chirp mass with orbital frequency colored. A green line representing the maximum spin-up to which continuous-wave searches have considered is also plotted \cite{abbott2019all}. Colored points below the green line, meaning smaller spin-ups, represent possible masses of inspiraling PBHs that can be probed with continuous-wave methods. The maximum chirp mass of a PBH merger that we could detect with continuous-wave methods is $\sim5\times 10^{-4}M_\odot$. 
    %corresponding to a gravitational-wave frequency of 40 Hz. 
    Transient continuous gravitational-wave methods are necessary to exhaustively constrain larger PBH chirp masses.}
    \label{fdotcw}
\end{figure}

% Continuous-wave searches can in principle detect PBHs given that their coalescence time is sufficiently long and the linear approximation does not fail (see section \ref{linvspow} for more details). However currently these searches only probe a very small portion of the PBH chirp mass parameter space. 
In Figure \ref{fdotcw}, we plot the spin-ups associated with certain chirp masses, with the gravitational-wave frequency colored. Superimposed on this plot is the maximum spin-up for which continuous-wave methods search \cite{abbott2019all}. Only for very small PBH masses ($<2\times 10^{-5}M_\odot$) can continuous-wave search results actually be used to place constraints on PBH chirp masses. Therefore, continuous-wave searches alone do not adequately cover the parameter space associated with PBH mergers, meaning transient continuous-wave methods are necessary to place more stringent constraints on a wider range of PBH chirp masses.

\section{primordial black hole \\ merging rates} \label{pbhmerge}

PBH mergers are a possible source of transient continuous waves and continuous waves. However, we wish to evaluate the rates at which these systems will merge to determine the chance we have to actually see an event. There are three possible formation channels of PBH binaries that would imply different merger rates, which are described in detail in sections \ref{primordial}, \ref{capture} and \ref{captureGC}. We then introduce the considered PBH mass distribution functions in section \ref{pbhsub}, which have a direct impact on the inspiral rates.
% {Our results are summarized in Figures~\ref{fig:rates} and~\ref{fig:sossys} displaying the expected rates in our galaxy, in the galactic center and in the solar system vicinity, for the different cases, as well as the expected number of sources at a given continuous wave frequency.} %% @ Sébastien, I don't want to 'forward reference' - reference figures out of order
%$f_{\rm gw}$, obtained through
%\be
%N^{\rm sources} (f_{\rm gw} ) = R /times t(f_{\rm gw})
%\be
%where $t$ is the time during which a binary emits gravitational waves in some frequency band, which is approximatively the time before they merge. }

\subsection{Primordial binaries} \label{primordial}

%\subsubsection{Origin}
Before matter-radiation equality in the early universe, if two PBHs formed sufficiently close to each other for their gravitational attraction to overcome the Hubble-Lema\^itre expansion of the Universe~\cite{Nakamura:1997sm,Sasaki:2016jop}, they form a binary 
%have been created and 
whose orbital characteristics depend on %sustained by 
the gravitational pull from other nearby PBHs. 
% Primordial binaries could have been created in the early Universe, before the matter-radiation equality, when two PBHs formed sufficiently close to each other for their gravitational attraction to overcome the Hubble-Lema\^itre expansion of the Universe~\cite{Nakamura:1997sm,Sasaki:2016jop}.  
% The gravitational influence of one or several nearby PBHs prevents the two black holes to immediately merge and instead they form a binary.  
Eventually, the binary is sufficiently stable and it takes of the order of the age of the Universe for the two black holes to merge. 
% \textcolor{red}{@Sebastien could you add some references for this process, and clarify if this happens independently of the PBH mass. {SC: I added 2 refs in the above paragraph. Merging rates depend on the PBH mass, as explained after, but the formation of such binaries is possible at any mass.}}  

%\subsubsection{Cosmological merging rate}

Assuming that the PBH spatial separation at formation is purely Poissonian, the cosmological merging rate today is approximately given by~\cite{raidal2019formation,Gow:2019pok,liu2019effects,Kocsis:2017yty}:
  %  $$\tau \sim  f_{\rm PBH} \times \mathcal 10^4) \, {\rm yr^{-1} \, Mpc}^{-3}~.$$
\bea
        R^{\rm cos}_{\rm prim} &\equiv & \frac{{\rm d}\tau}{{\rm d} \ln m_1 d \ln m_2} \nonumber \\ &\approx & \frac{1.6 \times 10^6}{\rm Gpc^3 yr} f_{\rm sup} f_{\rm PBH}^{{53/37}} \left(\frac{m_1 + m_2}{M_\odot}\right)^{-32/37} \nonumber \\ & \times & \left[\frac{m_1 m_2}{(m_1+m_2)^2}\right]^{-34/37} f(m_1) f(m_2)~,  \label{eq:cosmomerg}
\eea
    where $\tau$ is the rate per unit of logarithmic mass of the two binary black hole components $m_1$ and $m_2$, $f_{\rm PBH}$ is the dark matter density fraction made of PBHs
    %, $m_1$ and $m_2$ are the masses of the two binary black hole components 
    and $f(m)$ is the density distribution of PBHs normalized to one ($\int f(m) {\rm d} \ln m = 1$).  We have included a suppression factor $f_{\rm sup}$~\cite{Clesse:2020ghq} that effectively takes into account a rate suppression due to the gravitational influence of 
    %to the binary disruption
    % \textcolor{red}{@Sebastien what is binary disruption?} {SC: I have rephrased the first part of the sentence.} 
    early forming PBH clusters~\cite{raidal2019formation,Vaskonen:2019jpv}. %meaning that nearby PBHs tend to circularize the orbits of other PBH binaries, which relates to $t_{\rm merg}$. 
    The importance of this effect is still uncertain for general PBH mass functions.  We consider $0.001 \lesssim f_{\rm sup} \leq 1$ as a plausible range when $f_{\rm PBH} \gtrsim 0.1 $ and no suppression ($f_{\rm sup} \approx 1$) {when $0.005 \lesssim f_{\rm PBH} \lesssim 0.1 $}, which is motivated by $N$-body simulations ~\cite{raidal2019formation}\footnote{Nevertheless, when our analysis was almost terminated, a new study has claimed that the rates for primordial binaries are highly suppressed compared to previous calculations~\cite{Boehm:2020jwd}, an effect due to the variation of the Misner-Sharp mass as a function of the Universe expansion that is relevant to calculate the binary collapse time, as long as the binary does not belong to a virialized halo.  If these results are correct, only binary formation by tidal capture in dense PBH clusters would be relevant for our analysis.}.  {For lower values of $f_{\rm PBH}$, the suppression factor becomes lower than one (see e.g. Eq. 2.38 of~\cite{raidal2019formation}), but as shown later, continuous-wave methods are not sensitive enough to probe such low values of $f_{\rm PBH}$.  We also note that depending on the PBH mass distribution, the process of rate suppression can be more complex and efficient, and therefore dependent, on the binary component masses.  This dependence eventually leads to the dominance of the second PBH binary formation channel in clusters that we consider later.}

    %However, if PBHs contribute importantly to the dark matter, then  N-body simulations have shown that early-forming clusters somehow suppress this rate, eventually down the rates inferred by LIGO/Virgo for $f_{\rm PBH} \simeq 1$.  In the unsuppressed case, assuming $f_{\rm PBH} \times f(m) \simeq 0.01$ at planetary masses in order to pass the microlensing limits
  
For planetary-mass mergers, the astrophysical range of continuous-wave searches does not exceed tens of kiloparsecs (see Section~\ref{theosens}).  One therefore needs to compute the merging rate in the Milky Way.  For this purpose, we follow~\cite{Clesse:2016vqa} and have assumed an Einasto dark matter halo profile~\cite{Haud:1986yj}, 
 \be
 \rho_{\rm DM}^{\rm gal} (r) = \frac{\rho_{-2}}{{\rm e}^{2n_p \left[ (r/r_{-2})^{1/n_p}-1\right]}}, \label{rhodmgalr}
 \ee
 where $r$ is the distance to the galactic center, $r_{-2} \approx 20$ kpc is  the  radius  at  which  the  logarithmic  slope of  the  profile  equals $-2$ and  $\rho_{-2} \equiv \rho(r_{-2}) \approx 2 \times 10^{-3}  M_\odot {\rm pc}^{-3}$, and $n_p=4$ is a parameter of the profile.
 Other profiles (e.g. Navarro-Frenk-White - NFW) could have been considered but without significantly impacting our predictions. These binaries are formed before galaxy formation and so their merging rates only depend on their number density, which is because more black holes in a halo would facilitate the capture of a lone primordial one.  But binary formation depends only on the total number of black holes existing, not on the local PBH number density.
%  \textcolor{red}{@Sebastien, why is this true?} {I have rephrased these sentences because it was indeed not clear.}%, unlike  the tidal  capture channel discussed the next section.  
% Only counts the total PBH mass in a given volume \textcolor{red}{not a complete sentence}.  
With the profile in equation \eqref{rhodmgalr}, we can then compute the expected merging rate within our galactic halo, towards the galactic center or in the solar system vicinity, using the simple relation 
 \be
 R_{\rm prim}^{\rm gal/GC/sol} = \int R^{\rm cos}_{\rm prim} \times \left( \frac{\rho^{\rm gal}_{\rm DM} }{\bar \rho_{\rm DM} }\right) {\rm d} V~, \label{rategalgcsol}
 \ee
 where $\bar \rho_{\rm DM}$ is the mean cosmological dark matter density today.
 %the dark matter density at our location. 
 %($\rho_{\rm DM}=6.4\times 10^{-5}$ J/m$^3$).
% \textcolor{red}{@Sebastien why is Gpc$^3$ written next to $\rho_{\rm DM}$?} {SC: because we integrate over some volume, which needs to be expressed in Gpc$^3$.} 
For the galactic rate ($R_{\rm prim}^{\rm gal}$), we considered a volume of $50$ kpc radius;
for the galactic center rate ($R_{\rm prim}^{\rm GC}$), we considered the central region with a $0.1$ kpc radius that spans an arc of about ten degrees in the sky, similar to the directional sensitivity in targeted searches with LIGO/Virgo.  
For the solar system vicinity rate ($R_{\rm prim}^{\rm sol}$), i.e. at a distance of $8$ kpc from the galactic center, we assumed a constant dark matter density of $2 M_\odot {\rm pc}^{-3}$, consistent with the galactic dark matter profile, i.e. $3.3 \times 10^5$ larger than the cosmological dark matter density today.
% \textcolor{red}{@Sebastien why do you assume this larger dark matter density?} {{SC:  this is the density obtained from the galactic profile at the position of the solar system}.  
By integrating equation \ref{rategalgcsol} with the dark matter halo profile of equation \ref{rhodmgalr}, we obtain the following rates: 
% in terms of  $R_{\rm prim}^{\rm gal/GC/sol}$:
% \textcolor{red}{@Sebastien how do you obtain the following? Why is $R^{\rm cos}_{\rm prim}$ still a variable in the equations below if you integrate over it in eqn. 3?  } {SC:  Eq. 1 provides the cosmological rates.  But here we are interested by local scales, and so we need to consider the amount of dark matter in the galaxy, galactic center or nearby the solar system.}
\bea \label{tet}
 R_{\rm prim}^{\rm gal} & \approx &   2.2 \times 10^{-8} \,  \times R^{\rm cos}_{\rm prim},  \label{eq:galrates} \\
 R_{\rm prim}^{\rm GC} & \approx &  1.1 \times 10^{-13}  \,  \times  R^{\rm cos}_{\rm prim},   \label{eq:GCrates}   \\
 R_{\rm prim}^{\rm sol} & \approx &  1.4 \times 10^{-21} \,  \left( \frac{d}{\rm pc} \right)^3 \times  R^{\rm cos}_{\rm prim}~,  \label{eq:solrates} 
\eea
% \textcolor{red}{@Sebastien why is there a $d$ in equation 6 but not 5 or 4? Could you also express the numbers in a more realistic way and explain the meaning of equations 4-6 compared to 1 and 3? Also, do you make use of the mass functions to calculate these rates? If so, the rate calculations should be after the PBH mass functions section.} {{SC: Eq. 6 provides the rates in a sphere of radius $d$ around the solar system.  Eq 4 and 5 are for the whole galaxy and the galactic center (described as a sphere of 0.1 kpc as explained above), so $d$ is fixed by definition.  The mass function is hidden in $R_{prim}^{cos}$ so Eqs. 4-6 are general and do not yet assume a particular mass function.    }} 
where $d$ is the considered maximal distance of the PBH binary. Based on Equations \ref{tet}--\ref{eq:solrates} and the particular mass functions that will be described in section \ref{pbhsub}, planetary-mass PBH binaries can reach a yearly rate larger than one.  Furthermore, for small values of the chirp mass, one eventually probes lower orbital frequencies and the continuous waves are emitted long before the binary merger, which can last for many years in the same frequency band.  Therefore the number of primordial PBH binaries can be larger than one even if the corresponding merging rate is $R_{\rm prim} \ll 1 \, {\rm yr}
^{-1}$.  This number is approximately given by 
\be
N_{\rm prim}^{\rm gal/GC/sol} \approx R_{\rm prim}^{\rm gal/GC/sol} \times t_{\rm merg}~.
\ee 

     \subsection{Capture in primordial black hole halos}   \label{capture}

The second possible binary formation channel is through dynamical capture in dense PBH halos~\cite{Bird:2016dcv,Clesse:2016vqa}.  As any other dark matter candidate, PBHs are expected to form halos during cosmic history.  Their clustering properties determine the corresponding merging rate, which can be $\mathcal{O}(1-100) \, {\rm yr^{-1} \, Gpc}^{-3}$ for a standard Press-Schechter halo mass function~\cite{Bird:2016dcv}.  For realistic extended mass functions, the abundance, size and evolution of PBH clusters is impacted by several effects:  the Poissonian noise at formation, heavy PBH seeds, the shape of the primordial power spectrum, the dynamical heating of clusters, hierarchical mergers, cluster evaporation, etc.  These effects can  either boost or suppress the PBH merging rates, which are thus largely model-dependent.  Nevertheless, the mass dependence of the rates should remain a typical signature of mergers in clusters.  The cosmological merging rates are approximately given by~\cite{Clesse:2020ghq}
\bea
        R^{\rm cos}_{\rm capt} &\equiv & \frac{{\rm d}\tau}{{\rm d} \ln m_1 d \ln m_2} \nonumber \\ &\approx & R_{\rm clust.}  f_{\rm PBH} f(m_1) f(m_2) 
        %\nonumber \\  
        %& \times & 
        \frac{(m_1 + m_2)^{10/7}}{(m_1 m_2)^{5/7}},  \label{eq:cosmomergcapt}
\eea
where $R_{\rm clust.} $ is an effective scaling factor that incorporates the PBH clustering  properties.  As discussed above, the value of $R_{\rm clust.} $ is uncertain and model dependent, and yet LIGO/Virgo black hole merging rates can be explained if PBHs constitute a significant fraction of dark matter. Furthermore, for $f_{\rm PBH} = 1$, a value $R_{\rm clust.} \approx 400 {\rm yr^{-1} Gpc^{-3}} $  explains well the rates inferred from the recent events GW190425 and GW190814 with one suspected object in the black hole mass gap, from GW190521 with one object in the pair-instability mass gap, and for other events whose black hole masses were around $30 M_\odot$, while being consistent with LIGO/Virgo sub-solar rate limits~\cite{Clesse:2020ghq}.  We consider this value as our benchmark model.   {Such a value of $R_{\rm clust.}$ is also motivated theoretically when accounting for the enhanced clustering induced by Poisson fluctuations in the initial PBH spatial distribution.  This effect introduces a new term in the matter power spectrum, and according to the extended Press-Schechter formalism, the fraction of inhomogeneities that collapse into halos of mass below $10^7 M_\odot$ is close to unity for significant values of $f_{\rm PBH}$, as pointed out in~\cite{Kashlinsky:2016sdv} in a different context.  Combined with the low relaxation time of low-mass sub-halos, this sets a natural clustering scale for which one can naturally have $R_{\rm clust.} \sim [100-1000]$. }

In addition to calculating the merging rates for primordial binaries, we can use equations ~(\ref{eq:galrates}) and (\ref{eq:solrates}) for an estimation of the rates from PBH clusters in the entire galactic halo or in the vicinity of the solar system.  For planetary-mass PBH binaries with a mass ratio close to unity, these merging rates are a few orders of magnitudes lower than for primordial binaries.  However, these rates are boosted for very low mass ratios, for instance if the main black hole component has a stellar-mass while the secondary object has a planetary mass.  This boost is even more important if there is a peak in the PBH mass function at stellar masses, as expected from the thermal history of the Universe (see Section~\ref{pbhsub}).  For this binary formation channel, we will thus focus on the mass function accounting for the thermal history.

   \subsection{Capture in the Galactic Center}   \label{captureGC}
   
   Dense PBH clusters in the galactic center are probably unstable and tidally disrupted by the interactions with the central super-massive black hole and/or other clusters.  It is therefore expected that the PBH density distribution is smoother and follows the dark matter density profile, given by equation ~\ref{rhodmgalr}.  In the galactic center, one has 
   \be
   \rho_{\rm DM}^{\rm GC} \approx \rho_{-2} {\rm e}^{4n} \approx 6 M_\odot {\rm pc}^{-3} \approx 1.8 \times 10^8 \bar \rho_{\rm DM}~,
   \ee
    where $\bar \rho_{\rm DM} $ is the mean cosmological dark matter density today. 
    Let us nevertheless notice that the dark matter density in the galactic center depends on the shape and parameters of the profile.  The rate at which binaries form through tidal capture depends on the square of the local PBH density $R_{\rm capt}^{\rm GC} \propto (\rho_{\rm DM}^{\rm GC})^2$ and scales nonlinearly with the mean relative black hole velocity $R_{\rm capt}^{\rm GC} \propto  v_{\rm rel}^{-11/7}$.  We can therefore relate the galactic center rate to the cosmological rate of binary formation in clusters through:
    \be
    R_{\rm capt}^{\rm GC} \approx \frac{(\rho_{\rm DM}^{\rm GC})^2 V^{\rm GC}}{\rho_{\rm DM}^{\rm clust} \bar \rho_{\rm DM} {\rm Gpc^3} } \left( \frac{v_{\rm rel}^{\rm GC}}{v_{\rm rel}^{\rm clust}} \right)^{-11/7} \times R^{\rm cos}_{\rm capt},
    \ee
where $ \rho_{\rm DM}^{\rm clust} $ is the density of PBH clusters and $V^{\rm GC}$ is the considered volume in the galactic center.  A value of $R_{\rm clust} \approx 400 {\rm yr^{-1} Gpc^{-3}} $ corresponds to a cluster density of $\rho_{\rm DM}^{\rm clust}  = 6 \times 10^8 \bar \rho_{\rm DM} $ for a mean relative velocity of $v_{\rm rel}^{\rm clust} = 2 \, {\rm km/s}$.   The dynamics in the galactic center is a complex process, but the magnitudes of relative velocities should be of order $ v_{\rm rel}^{\rm GC} \sim 10 {\rm km/s}$, which leads us to an estimate the expected PBH capture rate:
\be
R_{\rm capt}^{\rm GC} \sim 3 \times 10^{-12} \times R_{\rm capt}^{\rm cos}.\label{rcapt}
\ee
This estimate is order-of-magnitude, given the different astrophysical uncertainties.  Equation \ref{rcapt} implies that the ratio between the galactic center and cosmological rates is roughly one order of magnitude larger for tidal capture than for primordial binary formation, which partially compensates for a lower value of $R^{\rm cos}_{\rm capt}$ compared to $R^{\rm cos}_{\rm prim}$.
 %\textcolor{red}{Is it not worth it to compute the rates with a realistic mass function?}

\subsection{Primordial black hole mass functions}
\label{pbhsub}

The distribution of PBH masses, a.k.a. the PBH mass function, remains largely unknown. We do not expect a monochromatic mass function (i.e. all PBHs with a single mass) because even in the limiting case of black hole formation due to a sharp peak in the primordial power spectrum, PBHs would acquire a wider distribution due to effects related to the critical collapse.  And for a broad mass function covering the planetary and the stellar-mass ranges, one expects features~\cite{Byrnes:2018clq,Carr:2019kxo} coming from the thermal history of the Universe and the transient variations of the equation of state that occur when the different species (e.g. the Higgs boson, top quark, W and Z bosons, protons, neutrons, and pions) become non-relativistic. Still, the exact shape  depends on the underlying primordial power spectrum.
   
For these reasons, we consider two cases for the PBH mass function:
    
 %   \begin{itemize}
  %      \item 
  \textit{Case 1 - Agnostic mass function:}  For a given PBH mass $m_{\rm PBH}$, we let $\tilde f(m_{\rm PBH}) \equiv f_{\rm PBH} f(m_{\rm PBH}) \sqrt{f_{\rm sup}}$ be a free model parameter and consider only almost equal-mass mergers that produce the highest strain, giving a merging rate"
\be 
%\tau(m_{\rm PBH})
R_{\rm prim}^{\rm cos} (m_{\rm PBH}) \approx \frac{1.7 \times 10^6}{\rm Gpc^3 yr} \tilde f(m_{\rm PBH}) \left(\frac{m_{\rm PBH}}{M_\odot}\right)^{-0.86}~.
\ee
% \textcolor{red}{@Sebastien How does this rate relate to the rates calculated in the previous section? Why do you calculate rates in the previous section and this one?}  {I have rephrased to make this more clear.  It is the cosmological merging rate when we have assumed equal-mass mergers.} 
The merging rate roughly scales as $1/m_{\rm PBH}$ and so one expects light PBHs to merge more often than heavy ones.  This case can represent sharp peaks in the power spectrum, or can be used to estimate an upper bound on the abundance at a given mass of PBHs with an arbitrary mass function.
        
\begin{figure}
 %   \centering
    \hspace{-3mm} \includegraphics[width=0.45\textwidth]{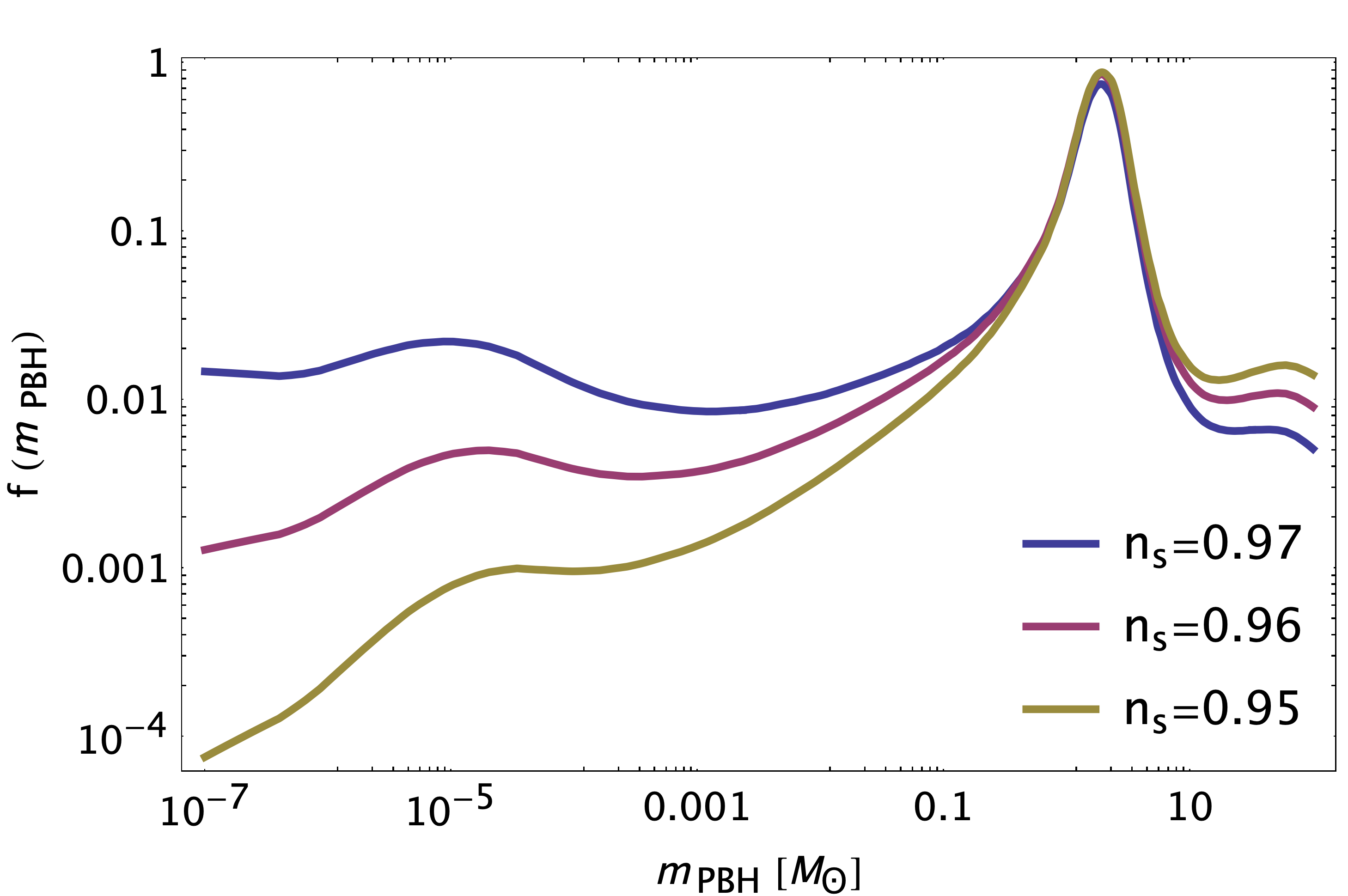}
    \caption{PBH mass distribution normalized to $f_{\rm PBH} = 1$, for Case 2. This model includes the equation of state reduction effects at the Quantum Chromodynamics transition on PBH formation for primordial power spectra with indices $n_{\rm s} = 0.95/0.96/0.97$ (yellow, red and blue curves).}
    \label{fig:fPBH}
\end{figure}

        %\item 
\textit{Case 2 - Thermal mass function:}  The primordial power spectrum of curvature fluctuations at the origin of PBH formation is nearly scale-invariant, with a spectral index $n_{\rm s} = 0.97$.  We include the features in the mass distribution due to thermal history and the progressive reduction of the number of relativistic degrees of freedom in the early Universe, following~\cite{Carr:2019kxo}.  The resulting PBH mass function is displayed in Figure~\ref{fig:fPBH}.  This spectral index is compatible with observational limits and could give evidence of PBHs in LIGO/Virgo observations, in particular the masses and rates of the recent mergers GW190425 and GW190814 with $f_{\rm PBH} = 1$ (and $f_{\rm sup} \sim 0.01$)~\cite{Clesse:2020ghq}.   Moreover, a value  $n_{\rm s} \gtrsim 0.98$ or $n_{\rm s} \lesssim 0.95$ would lead to an overproduction of light or heavy black holes respectively. Using the full mass distribution, we have computed the expected merging rate as a function of the binary chirp mass, ${\rm d} \tau / {\rm d} \log \mathcal{M}$, which does not only include equal-mass binaries but also binaries with lower mass ratios.  The limits on this scenario are set on the parameter combination $\bar f_{\rm PBH} \equiv f_{\rm PBH} \sqrt{f_{\rm sup}}$.

\begin{figure*}[ht!]
    \centering
    \hspace{-3mm} \includegraphics[width=0.50\textwidth]{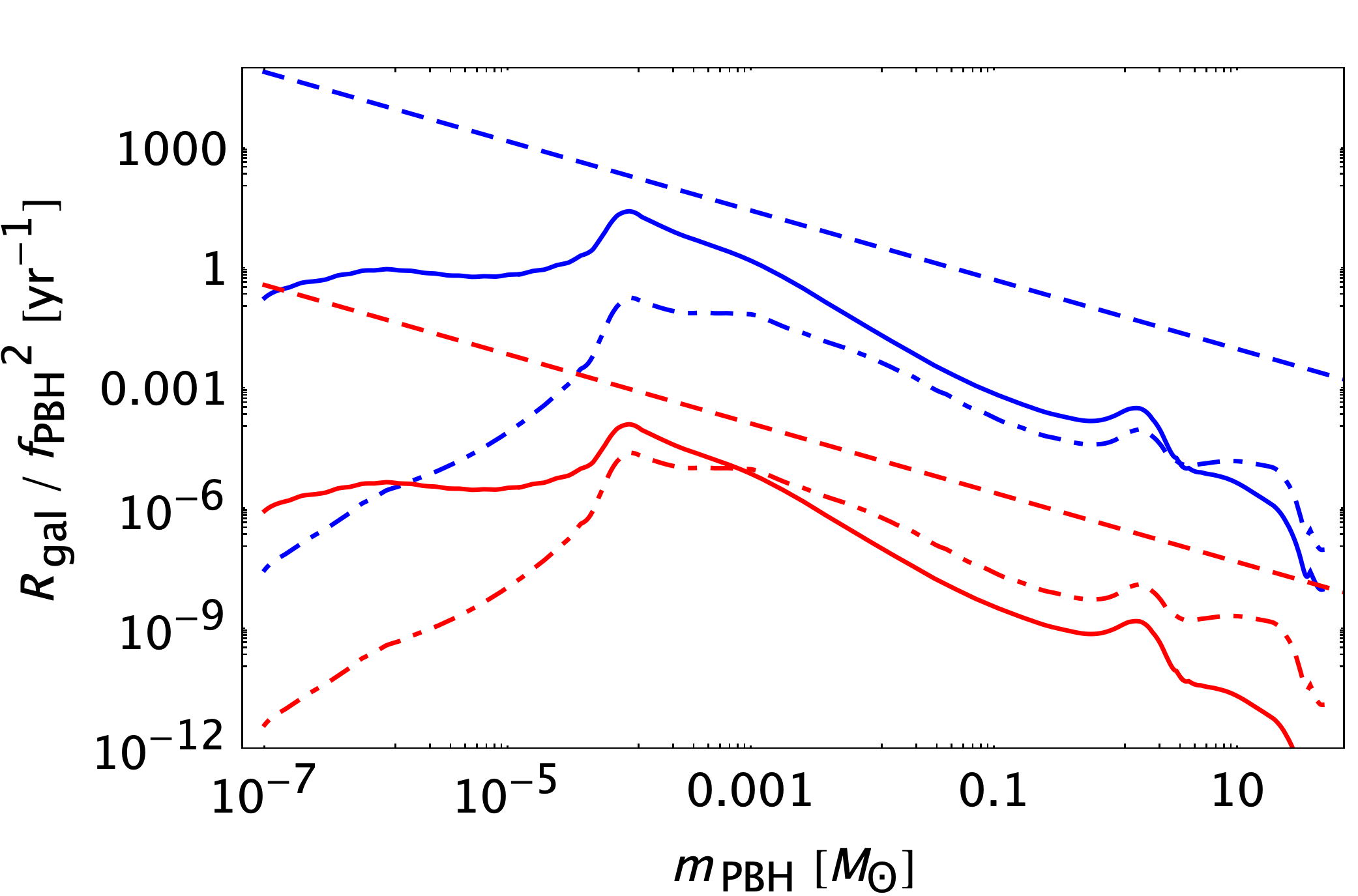}%rates_vs_mass.png}
    \includegraphics[width=0.50\textwidth]{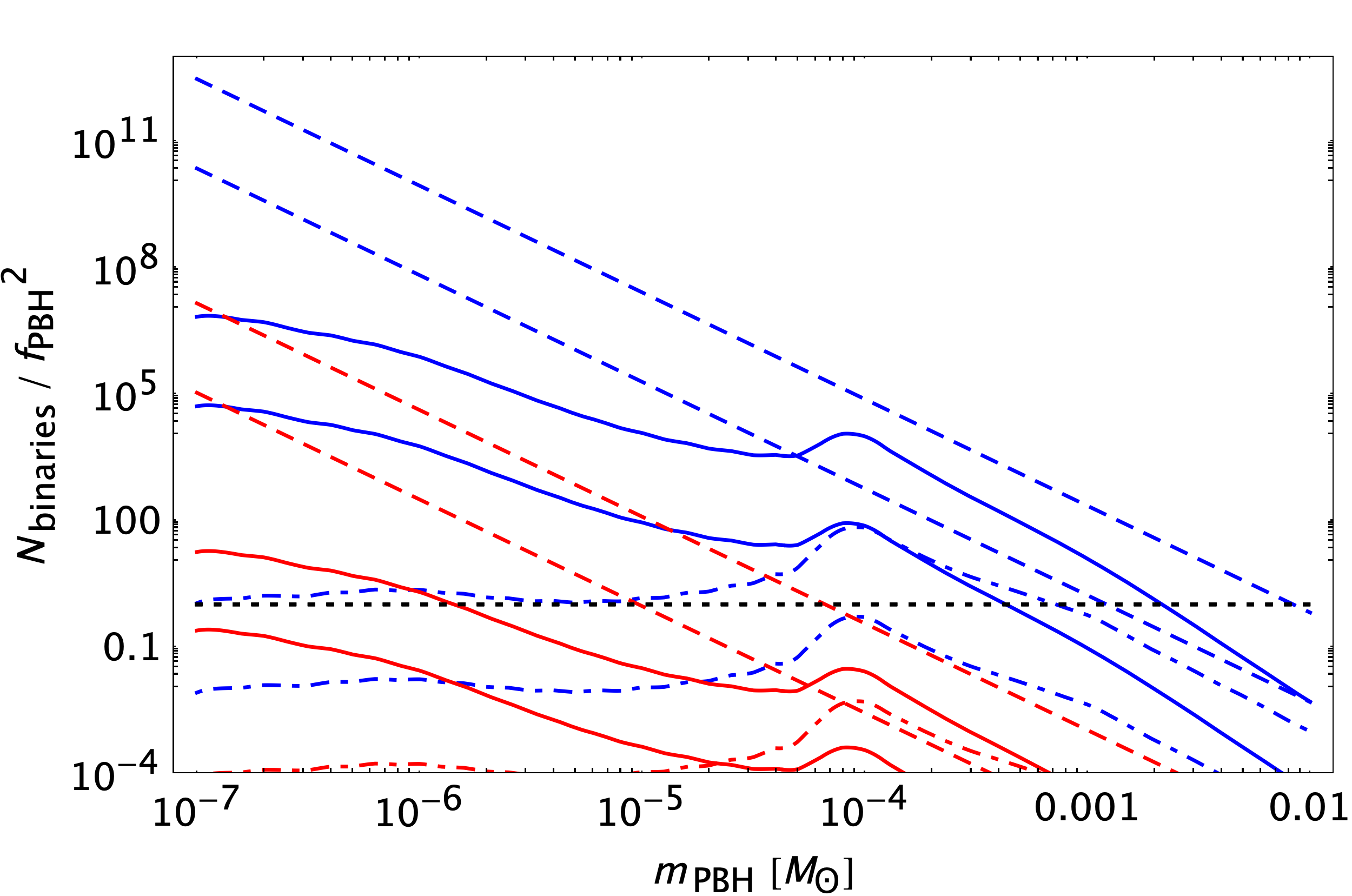}%Nbins_vs_mass.png}
    \caption{{Left:  Merging rates in the galactic halo (blue lines) and galactic center (red lines) for $f_{\rm PBH} = 1$,
    %and in our solar system (green lines, distance less than 10 pcs) 
    %divided by $f_{\rm PBH}^2$, 
    from primordial binaries in Case 1 (dashed lines) and Case 2 (solid lines), and tidal capture in Case 2 (dotted-dashed lines).  Right:  Expected number of binaries, for the same models and cases, for two typical gravitational-wave frequencies, $10$ Hz and $50$ Hz, corresponding to the maximal astrophysical range of Einstein Telescope (upper lines) and LIGO/Virgo (lower lines).  The rates and number of sources for different PBH fractions are obtained by rescaling by $f_{\rm PBH}^2$.} 
    %two choices of the spectral index. 
    %\textcolor{red}{@Sebastien why are there two of each curve now in the right panel of figure 3? Different spectral indices or something else? {SC: it is for case 1 (agnostic mass function) and case 2 (thermal mass function), now included in the caption.}} \textcolor{red}{No, I mean why are there 2 solid blue lines? and 2 lines of every type for every color, for a total of 12 lines etc. , compared to the left panel, where there are a total of 6 lines.}
    % \textcolor{red}{@Sebastien can you refer to this figure in the main text? Also: why are these rates divided by $f_{\rm PBH}^2$? Does this division mean that the rates are independent of how much dark matter makes up the universe?}  {The figures shows the rate in the case 100\% of dark matter is made of PBHs.  If one wants the rates for a lower dark matter fraction, he can just rescale the curves by multiplying by $f_{PBH}^2$}.  
    }
    \label{fig:rates}
\end{figure*}

For both agnostic and thermal PBH mass functions, we plot the merging rates and the number of expected binaries as a function of the chirp mass of a PBH binary system in left- and right-hand panels of figure \ref{fig:rates}, respectively. These curves assume that dark matter is entirely composed of PBHs ($f_{\rm PBH}=1$), though they can be rescaled for other assumptions of this fraction simply by multiplying by $f_{\rm PBH}^2$. The rates for PBH binaries with lower chirp masses are higher compared to those for higher chirp masses assuming an agnostic mass function, but exhibit a small peak around $10^{-4}M_\odot$ when using the thermal mass function.

\begin{figure}[ht!]
    \centering
    \hspace{-3mm} \includegraphics[width=0.45\textwidth]{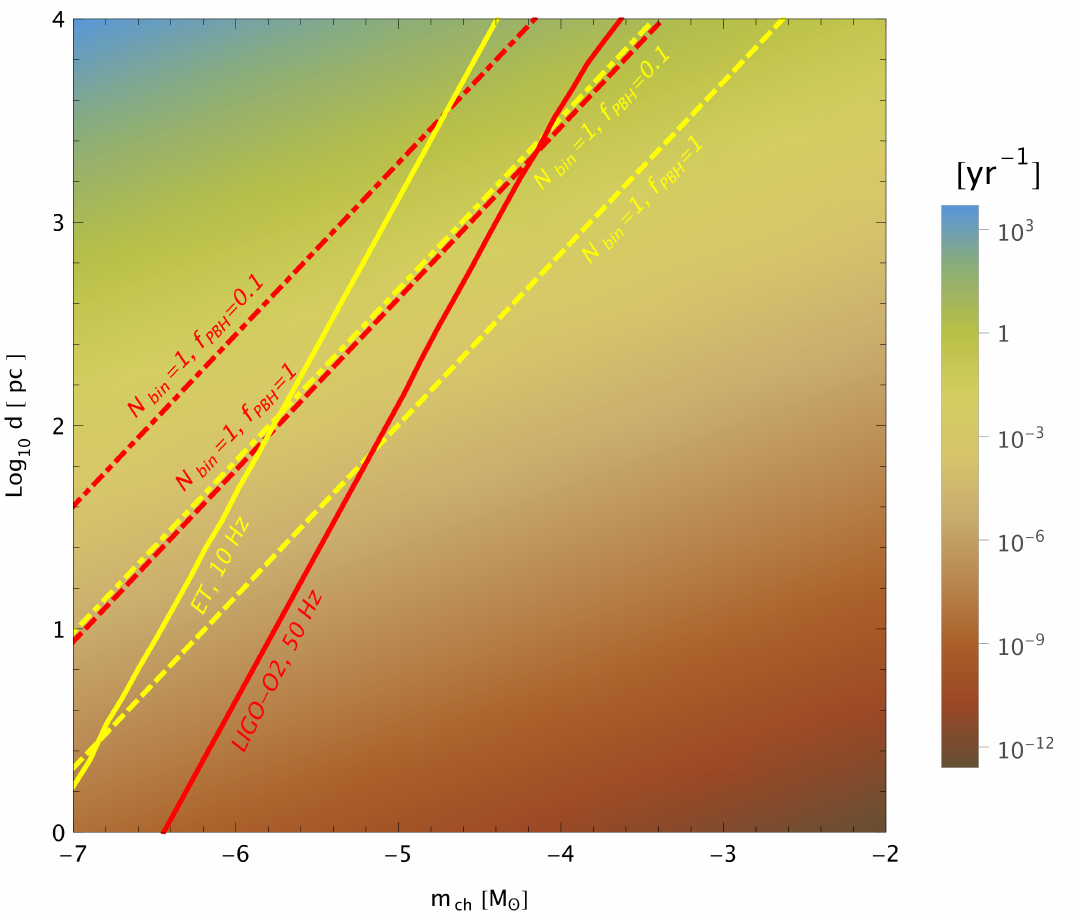}
    \caption{PBH merging rates within a sphere of radius $d$ centered on the solar system, for Case 1 with $f_{\rm PBH} = 1$. The colored solid lines represent LIGO/Virgo-O2 (red) and Einstein Telescope (ET) sensitivities at 50 Hz and 10 Hz, respectively, for the analyses methods described later in section \ref{theosens}. Dashed and dash-dotted lines correspond to one binary inspiral at $f_{\rm PBH} = 1$ and $f_{\rm PBH} = 0.1$, respectively, for LIGO/Virgo-O2 (red) at 50 Hz and ET (yellow) at 10 Hz. When the solid curves go above the dashed or dash-dotted lines, we can claim that one binary inspiral at a particular chirp mass and distance could be detected by LIGO/Virgo or Einstein Telescope. }
    % \caption{PBH merging rates within a sphere of radius $d$ centered on the solar system, for Case 1 with $f_{\rm PBH} = 1$.  The colored oblique lines correspond to a number of binary $N_{\rm bin} = 1$ (solid) and $0.1$ (dashed), at frequency above $8$ Hz (yellow lines)) and $50$ Hz (red lines). 
    % \textcolor{red}{@Sebastien can you please refer to this figure in the main body of the paper with some more explanation?}}
% cd ..    
    \label{fig:sossys}
\end{figure}

It is worth understanding whether the rates of PBH mergers at particular distances are detectable with current and future gravitational-wave detectors. Figure \ref{fig:sossys} shows merger rates (colored) as a function of the distance reach and the PBH chirp mass. Superimposed on this plot are the solid lines that give the sensitivities, calculated in section \ref{theosens}, of LIGO/Virgo and Einstein Telescope at particular frequencies to these mergers, and dashed lines that give the distance at which we would be able to detect one PBH inspiraling binary assuming that PBHs compose 10\% or 100\% of dark matter. When the solid line is above the dashed line (of the same color), we are able to detect an inspiraling PBH signal. Our analysis method will be explained in the following sections of the paper.

 %   \end{itemize}

\section{Search methods}\label{searchmethod}

Though there are many different formation channels for PBHs, our data analysis methods are not explicitly sensitive to them. Rather, we first endeavor to find a power-law or quasi-monochromatic signals from a particular inspiral, and use that information to derive the rates and to forecast limits on the fraction of dark matter that PBHs could compose. In this way, our analysis is independent of how PBHs form, though in the case of a detection, our results can be interpreted in the context of a wide variety of formation models based on different mass functions.

% The analysis begins with the construction of time/frequency ``peakmaps'', which are the inputs to both methods described in this section.

The analysis begins with time-domain strain data. From this data, we take Fast Fourier Transforms of different durations, $\TFFT$ \cite{sfdb_paper,piccinibsd}, estimate the background using an auto-regressive method, equalize the power spectrum and select local maxima above a certain threshold $\theta_{\rm thr}=2.5$, chosen as a compromise between sensitivity and computational cost \cite{sfdb_paper}. This process allows us to form a time/frequency ``peakmap'', as shown in the left-hand plot of figure \ref{pm_hm_inj}. The color represents the equalized power spectra at each time in a particular frequency band. There is a very strong simulated inspiral signal due to an injected inspiraling PBH binary system. These peakmaps are the input to the two methods described in this section.
% We can view the peakmap as a collection of ones and zeros: ones where there is color, and zeros elsewhere. In fact, the following methods do not make use of the strength of the signal, or the color, of the peakmap at all, but just whether at a given time/frequency, there is a peak or not. In this way, the methods avoid being blinded by noise artefacts, and gaps in the detector data.
One method is the Frequency-Hough transform that has been designed to detect signals whose frequency evolution is linear, discussed in section \ref{fht}; the other is the Generalized Frequency-Hough transform, which is conceptually similar to the Frequency-Hough transform, but can detect any signal whose frequency evolution follows a power law, which is explained in section \ref{ght}. Finally in section \ref{pproc}, we describe the post-processing steps taken after the Frequency-Hough or Generalized Frequency-Hough is run, which involve selecting significant candidates in each detector and performing coincidences between them.

\subsection{Frequency-Hough Transform}\label{fht}

The Frequency-Hough transform \cite{Astone:2014esa} searches for continuous waves from asymmetrically rotating neutron stars by mapping points in the detector time/frequency plane to lines in the source's frequency/spin-down plane. This transformation is done for each sky position. First, the time/frequency peakmap is corrected for the relative motion of the earth and source, and then the Frequency-Hough is performed. The signal model is a Taylor series expansion of the frequency in time, neglecting nonlinear terms:

\begin{equation}
    f=f_0+\dot{f}(t-t_0),
    \label{taylor}
\end{equation}
where $f$ is the frequency at time $t$ in the input time/frequency map, and $f_0$ is the intrinsic source frequency at time $t_0$. As long as the second order spin-down parameter $\ddot{f}$ is small and the observation time is not too long, this term can be safely neglected in continuous-wave searches, though recently $\ddot{f}$ has been used in a directed search \cite{aasi2015searches}. The Frequency-Hough has proven to be very sensitive to potential continuous-wave sources anywhere in the sky \cite{abbott2019all} and towards the galactic center \cite{piccinni2020directed}. 

\subsection{Generalized Frequency-Hough Transform}\label{ght}

Recently the Frequency-Hough was adapted to search for signals that last $\mathcal{O}$(hours-days), in which a power-law model for the frequency evolution in time is assumed \cite{PhysRevD.98.102004}, as opposed to a linear one:

\begin{equation}
\dot{f}=-k f^n,
\label{gfhpowlaw}
\end{equation}
and its integral over time is:

\begin{equation}
f(t)=\frac{f_0}{\left(1+k (n-1)f_0^{n-1}(t-t_0)\right)^{\frac{1}{n-1}}}.
\label{int_powlaws}
\end{equation}
Adapting the Frequency-Hough to search for power-law signals amounts to transforming the frequencies in the input time/frequency map first in the following way:

\begin{equation}
x=f^{1-n}. \label{xx}
\end{equation}
Once we have changed coordinates (by substituting equation \ref{xx} into equation \ref{int_powlaws}), the signal's frequency evolution becomes linear in the new space:

\begin{equation}
x=x_0+k(n-1)(t-t_0),
\end{equation}
where we have also written $x_0=f_0^{1-n}$. Now, points in the time/$x$ plane are mapped to lines in the $x_0/k$ plane, and these two variables translate directly back to $f_0$ and $\dot{f}_0$, and therefore $\mathcal{M}$. For inspiraling PBHs, these parameters allow us to calculate the chirp mass and the gravitational-wave frequency at a particular time before coalescence.

Note that to transition from searches for neutron stars spinning down to inspiraling PBHs spinning up, the only modification necessary is to allow $k\rightarrow -k$ in equation \ref{gfhpowlaw}. 

The Generalized Frequency-Hough has been developed and applied to a search for a long-lived remnant of GW170817 \cite{longpmr} and as a follow-up tool to a machine learning-based search for the same system \cite{miller2019effective}. Though the signal durations analyzed are shorter than canonical continuous waves, the Generalized Frequency-Hough requires a large amount of computational power because the number of points in the grid in $k$ becomes large in certain portions of the parameter space. These limitations will be discussed further in section \ref{anamethod}.

\begin{figure*}[ht!]
     \begin{center}
        \subfigure[ ]{%
            \label{pm_injt}
            \includegraphics[width=0.5\textwidth]{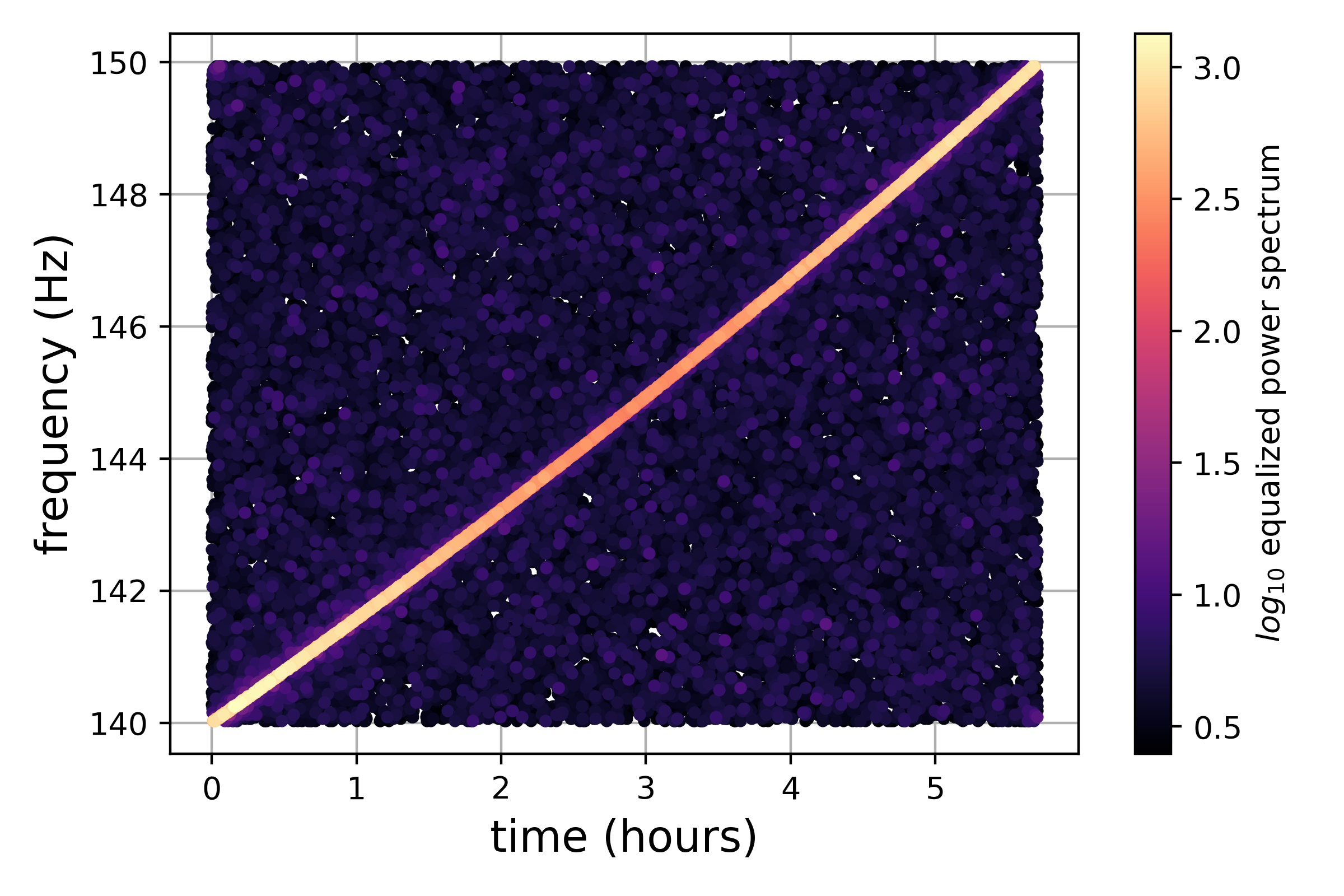}
        }%
        \subfigure[]{%
           \label{hm_inj}
           \includegraphics[width=0.5\textwidth]{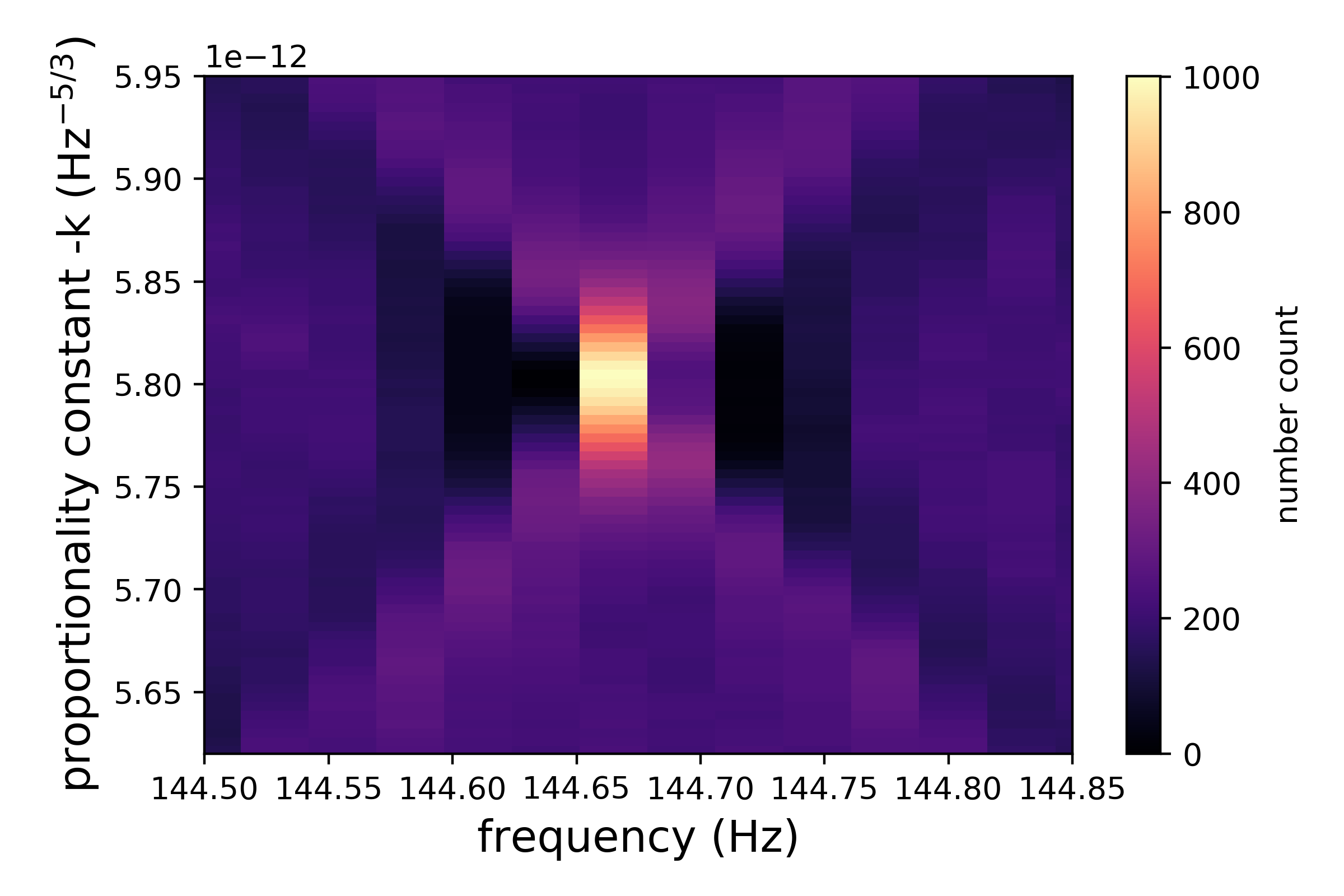}
        }\\ %  ------- End of the first row ----------------------%
%        \subfigure[Transformation of figure \ref{fig:second} to different coordinates]{%
%            \label{fig:third}
%            \includegraphics[width=0.49\textwidth]{figures/time_x_map_sig.jpg}
%        }%
%        \subfigure[Histogram showing injected signal's parameters]{%
%            \label{fig:fourth}
%            \includegraphics[width=0.49\textwidth]{figures/x0_k_hough_map_sig.jpg}
%        }\\ %  ------- End of the second row ----------------------%
%
%        \subfigure[Phase-corrected time/frequency map using \ref{fig:fourth}]{%
%            \label{fig:fifth}
%            \includegraphics[width=0.49\textwidth]{figures/time_freq_map_sig_phase_corr.jpg}
%        }%
%%
%        \subfigure[Hough map, phase-corrected signal's parameters]{%
%            \label{fig:six}
%            \includegraphics[width=0.49\textwidth]{figures/f0_fdot0_hough_map_ph_cor.jpg}
%        }%
%
    \end{center}
    \caption[]{%
   The left-hand plot shows the peakmap (time/frequency map), created with $\TFFT=\tfftused$ s, of a strong, rapidly evolving signal, which is the input to the Generalized Frequency-Hough transform. The right-hand plot shows the output of the Generalized Frequency-Hough transform, which is a histogram in the $f_0$/$k$ space of the source. The injection parameters are $h_0=\round \orighnaught$, $f_0=\round \adjfnaught$ Hz, $x_0=\roundfour\adjxnaught$ Hz$^{-8/3}$, 
    $\mathcal{M}=10^{-3} M_\odot$, and $k=\round \origk$ Hz$^{-5/3}$. The recovered candidate is in the same bin as the injection.} %$\roundtwonosci \binsaway$ bins from the injection.}%
   \label{pm_hm_inj}
\end{figure*}

% \subsubsection{Injection recovery}
We demonstrate that the Generalized Frequency-Hough transform can recover a signal from two inspiraling PBHs. A signal is injected in $\text{\detnameinj}$ data with parameters $f_0=\round \adjfnaught$ Hz 
% at $t_0=\round \adjtnaught$ MJD 
with $\mathcal{M}= \roundtwo \origmc M_\odot$ for a duration of $\dur$ s. The peakmap and its Generalized Frequency-Hough Transform are shown in left- and right-hand panels of figure \ref{pm_hm_inj}, respectively. After the Generalized Frequency-Hough transform, the signal is well-localized to one $x_0$ and $k$ bin. To create the peakmap, we used $\TFFT=\tfftused$ s.

%$t_0=\round \candtnaught$ MJD,

\subsection{Post-processing}\label{pproc}

We summarize how we determine significant candidates, which is done in the same way as in \cite{Astone:2014esa,PhysRevD.98.102004}.
After we run either the Frequency-Hough or Generalized Frequency-Hough transforms, we calculate a detection statistic called the critical ratio (CR), which depends on the number count $y$ in a particular $x/k$ bin and the mean $\mu$ and standard deviation $\sigma$ of the Hough map:

\begin{equation}
    CR=\frac{y-\mu}{\sigma}.
\end{equation}
On the basis of the number count and CR, significant candidates are selected uniformly in $f$ and $k$. This process is repeated for each detector in the network. Then, coincidences are done between the significant candidates returned by each detector. If the candidates' Euclidean distance in the $x/k$ plane is less than three bins away, the candidates are in coincidence and require further study. This coincidence distance $d_{\rm coin}$ is defined as:

\begin{equation}
    d_{\rm coin}=\sqrt{\frac{|x_2-x_1|}{\delta x}+\frac{|k_2-k_1|}{\delta k}}, \label{dcoin}
\end{equation}
where $x_1$/ $x_2$ and $k_1$/$k_2$ are the candidate's parameters returned by the analysis of the data from detectors 1 and 2, and $\delta x$ and $\delta k$ are the bin sizes in $x$ and $k$, respectively. 

To deeply analyze these candidates, we perform the follow-up, described extensively in \cite{Astone:2014esa,PhysRevD.98.102004}. From the initial analysis we have an estimation of the frequency at a reference time, and the chirp mass. With these two parameters, we know (almost) exactly the frequency evolution of the signal given in equation \eqref{powlaws}. We can then correct for the signal's phase evolution in the time domain, effectively making it monochromatic, and then take longer $\TFFT$ to determine if the signal's CR increases. If a coincident candidate does not follow this behavior, we veto it. 
% In practice, we consider a small window around the $x/k$ parameters to allow for the uncertainty {(SC: not clear what this means)} due to the discretization of the parameter space.

\section{Analysis of search methods}\label{anamethod}

The methods presented in section \ref{searchmethod} have been applied and tested extensively on canonical continuous-wave and transient continuous-wave signals from isolated neutron stars. Though the frequency evolution of the signal from PBH mergers is governed by the same equations as that for neutron stars, the analysis scheme has to be different, primarily because the system reaches its maximum spin-up at the \emph{end} of its life, not the beginning, in contrast to spinning down neutron stars. Our methods also work in two different regimes: one looks for quasi-monochromatic frequencies, the other looks for power laws. Understanding the ideal technique to use as a function of the parameter space is important to ensure optimal sensitivity towards PBH inspirals. We therefore explain in section \ref{linpl} the regions of the source parameter space for which we can assume the signal's frequency evolution is linear, for different choices of search parameters. In section \ref{ctobs}, we derive the optimal choice of $\TFFT$ and the time to observe $\Tobs$ as a function of the source parameter space, assuming that the source/earth Doppler effect can be corrected for perfectly. We then explore the computing costs for directed and all-sky searches for PBH mergers in section \ref{compcost}. Finally we derive an expression for the theoretical sensitivity of this search in section \ref{theosens}.

\subsection{Linear approximation versus power law}\label{linpl}
\label{linvspow}
Because continuous-wave signals are quasi-monochromatic and quasi-infinite, we can integrate over years of data to dig deeply into in the noise and provide strong constraints on the presence of asymmetrically rotating neutron stars in our galaxy \cite{abbott2019all}. However, the methods to search for power laws are not nearly as sensitive as continuous-wave methods, since the Fast Fourier Transform lengths they can take are limited by significantly higher spin-downs ($\mathcal{O}(10^{-3})$ Hz/s vs. $\mathcal{O}(10^{-9})$ Hz/s), and the signal lasts for a shorter duration (hours/days vs. years). It is therefore worthwhile to ask: for which observation times is the spin-up in equation \ref{chirp_pl} essentially constant, meaning that the frequency evolution is linear in time? In this way, we would be able to combine continuous-wave and transient continuous-wave techniques and cover the full parameter space with the optimal sensitivity. Here, we define that the linear approximation succeeds when the difference in frequencies calculated in equations \ref{taylor} and \ref{powlaws} are within one frequency bin $\delta f=1/\TFFT$.

\begin{figure*}[ht!]
     \begin{center}
        \subfigure[ ]{%
            \includegraphics[width=0.8\columnwidth]{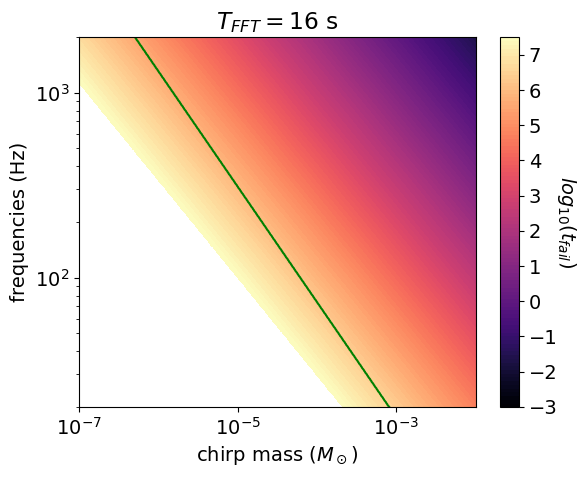}
        }%
        \subfigure[]{%
           \label{tfft16j}
           \includegraphics[width=0.8\columnwidth]{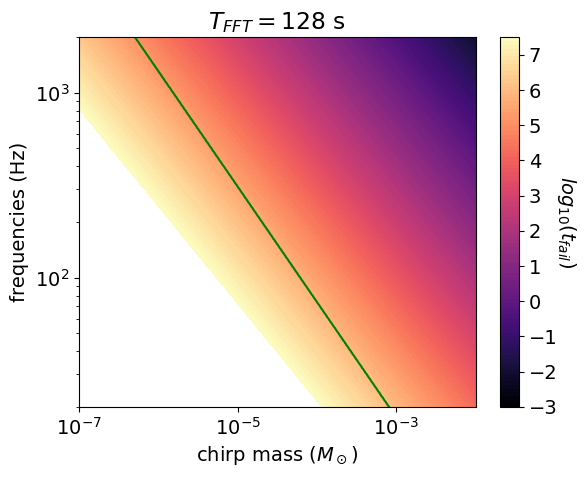}
        }\\ %  ------- End of the first row ----------------------%
        \subfigure[]{%
            \label{tfft128}
            \includegraphics[width=0.8\columnwidth]{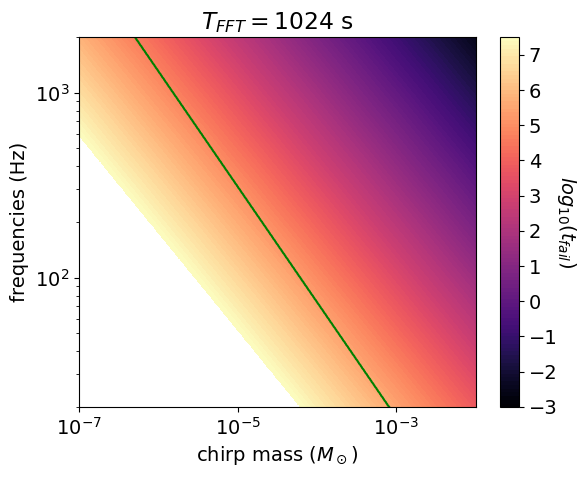}
        }%
        \subfigure[]{%
            \label{tfft8192}
            \includegraphics[width=0.8\columnwidth]{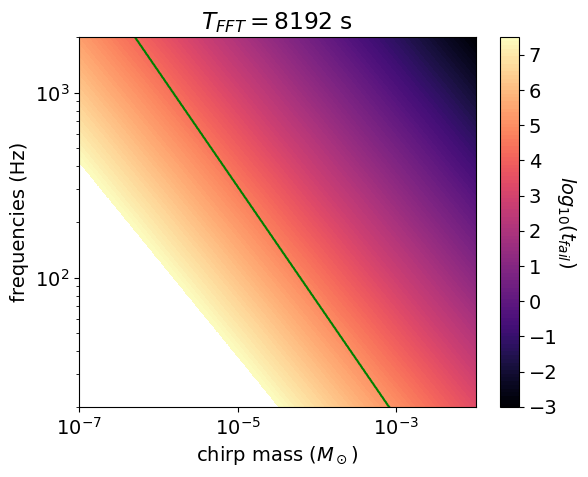}
        }\\ %  ------- End of the second row ----------------------%

%        \subfigure[Phase-corrected time/frequency map using \ref{fig:fourth}]{%
%            \label{fig:fifth}
%            \includegraphics[width=0.49\textwidth]{figures/time_freq_map_sig_phase_corr.jpg}
%        }%
%    %
%        \subfigure[Hough map, phase-corrected signal's parameters]{%
%            \label{fig:six}
%            \includegraphics[width=0.49\textwidth]{figures/f0_fdot0_hough_map_ph_cor.jpg}
%        }%
    \end{center}
    \caption[]{%
   This figure shows the maximum time that we can observe for to detect a PBH inspiral as a function of the initial frequency and chirp mass of the system, assuming the linear approximation in equation \ref{taylor} holds. These plots assume a maximum observation time of one year and $T_{\rm FFT}=16,128,1024,8192$ s, which represent typical transient continuous-wave and continuous-wave search choices. Above the green line on each plot represents the parameter space for which the observation time exceeds the time to coalescence, while white space indicates parameters for which the linear approximation never fails during the observation time. As $\TFFT$ increases, the linear approximation fails for more points in the parameter space (corresponding to less white space) because the frequency bin is smaller.
     }%
   \label{fail_lin_approx}
\end{figure*}

Figure \ref{fail_lin_approx} shows, for four different Fast Fourier Transform lengths $\TFFT$ assuming a maximum observation time of one year, the times at which the linear approximation fails, $t_{\rm fail}$, in the gravitational-wave frequency/chirp mass parameter space. Typical continuous-wave searches employ $\TFFT=1024-8192$ s \cite{abbott2019all}, while searches for transient continuous waves use $\TFFT=1-16$ s \cite{longpmr}. It is clear that a significant portion of the parameter space is not covered by continuous-wave searches, motivating the use of other methods to look for PBHs. White space in each plot corresponds to parameters for which the linear approximation never fails during the maximum observation time of one year. Above the green line corresponds to points in the parameter space for which $\Tobs$ exceeds the time to coalescence. The linear approximation fails for more portions of the parameter space for greater $\TFFT$ because the corresponding frequency bin is smaller. At smaller chirp masses, more of the parameter space is accessible to continuous-wave methods because of the small-enough spin-up during the observation time. 

\subsection{Cutoff time to observe}\label{ctobs}

In targeted continuous-wave searches, it is best to observe for as long as possible so that we can take longer and longer Fast Fourier Transforms, further isolating the signal and reducing noise to one smaller and smaller frequency bin. In transient continuous-wave searches, the signal does not last forever, and the spin-downs we consider are typically orders of magnitude higher than those in continuous-wave searches, which limits the $\TFFT$ we can take and therefore our sensitivity. 

\begin{figure}
    \centering
    \includegraphics[width=0.85\columnwidth]{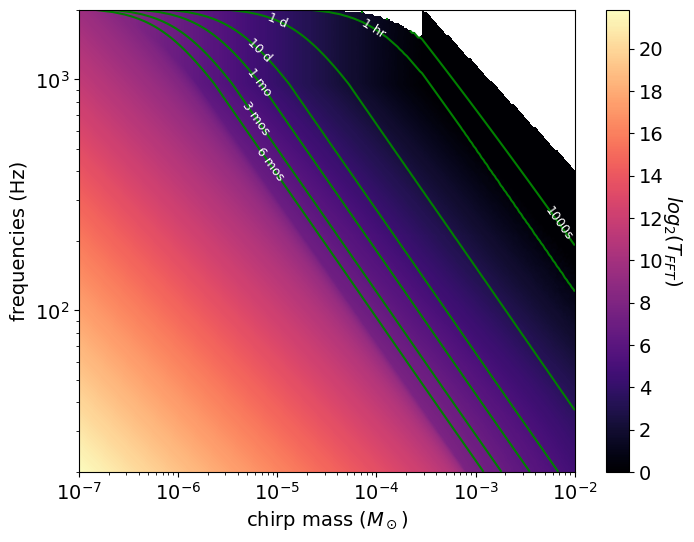}
    \caption{Source parameter space with optimal $\TFFT$ colored for particular observation times drawn as green contour lines. We impose that the signal cannot spin-up by more than 1 Hz/s and cannot reach a frequency greater than 2048 Hz. The optimal choices for $\TFFT$ and $\Tobs$ do not behave linearly at high frequencies, emphasizing the need for optimizing the choice of these two parameters. Moreover it is clear that for lower frequencies and lower chirp masses, we should observe for longer with higher $\TFFT$, which is consistent with the fact that signals in that portion of the parameter space are more similar to continuous waves than transient continuous waves. To create this plot, we use the advanced LIGO/Virgo sensitivity curves to determine $S_n$, and evaluate $S$ for a range of $\Tobs$ and $\TFFT$ at each point $f_0$ and $\mathcal{M}$ in the parameter space, and plot the combination of $\TFFT$ and $\Tobs$ that maximizes $S$. }
    \label{bestfft}
\end{figure}

For PBH signals, the spin-up increases with time as seen in equation \ref{chirp_pl}, which means that the corresponding $\TFFT$ we can take decreases, since we choose $\TFFT$ such that the power due to a signal with a particular spin-up $\dot{f}$ is confined to one frequency bin: $\dot{f}\TFFT<1/(2\TFFT)\rightarrow \TFFT<1/\sqrt{2\dot{f}}$. If we think about the sensitivity $S$ of a search in the following way \cite{PhysRevD.100.064013}:
\begin{equation}
S \propto \frac{h(f)}{\sqrt{S_n(f)}}{T^{1/4}_{\rm obs}}{T^{1/4}_{FFT}},
\end{equation}
there should exist optimal $\Tobs$ and $\TFFT$ based on the frequency evolution of the signal and the noise power spectral density of the detector $S_n(f)$ that maximizes $S$. More concretely, if we imagine a PBH system with $f_0=20$ Hz, the system will spin up into better and better frequency bands (lower values of $S_n$), gain in amplitude (higher values of $h(f)$) over more and more time (greater $\Tobs$) but with smaller $\TFFT$. And eventually, the signal will cross the bucket region in LIGO/Virgo and actually enter bands in which the detector sensitivity becomes worse. Therefore, observing for long periods when $\TFFT$ is small may not give us a gain in sensitivity. We investigated the interplay between these parameters for the mass range of PBH systems we consider, and determined, for each set of parameters, the optimal $\TFFT$ and $\Tobs$ to use in a search. These choices are shown in figure \ref{bestfft}, fixing a maximum observation time of one year. For small chirp mass at low frequencies, the signal is essentially monochromatic and close to a fully coherent analysis can be performed (bottom left-hand corner of figure \ref{bestfft}) if computational cost were not a concern. However, for most of the parameter space, semi-coherent analyses such as the Frequency-Hough and Generalized Frequency-Hough are necessary. Additionally, we see in the top-right hand corner that it is better to cut off the observation time at, say, one hour and use $\TFFT\sim 20$ s, than to continue to analyze data for longer periods of time, which would require a smaller $\TFFT$.

\subsection{Computational cost and search design}\label{compcost}

% We estimate the computation time for all-sky and directed searches for PBH inspirals.
There are two major contributors to the computational load of this search: (1) the grid we construct in $k$, corresponding essentially to a grid in chirp mass, and (2) the grid we construct in the sky. The first point is relevant only for signals for which the linear approximation fails, but the second one is important for both signals that follow both equation \ref{powlaws} and equation \ref{taylor}. We perform an estimation for a directed search 
% (one in which we assume to know the exact position of the source, e.g. pointing towards the galactic center) 
and an all-sky search.
% (one in which we must analyze each sky location individually).

\subsubsection{Directed search}\label{dirser}

For a directed search, for each combination of $f_0$ and $\mathcal{M}$, we extract a frequency band whose width is determined by the signal evolution during the observation time: $f_{\rm min}=f_0$; $f_{\rm max}=f(\Tobs)$, and create a peakmap. Each point in the parameter space $f_{\rm min},f_{\rm max},\TFFT, \Tobs$ determines a grid in $k$, which relates to the chirp mass. This grid in $k$ is constructed such that the power due to an inspiraling binary is confined to one $x$/$k$ bin when moving from $\mathcal{M}$ to $\mathcal{M}+\delta \mathcal{M}$, where $\delta \mathcal{M}$ is a small change in the chirp mass that relates to the spacing in $k$. See equations 21-24 in \cite{PhysRevD.98.102004} for more details on this grid. 

\begin{figure}[ht!]
    %  \begin{center}
%
        % \subfigure[ ]{%
            \label{compdirparm}
            \includegraphics[width=\columnwidth]{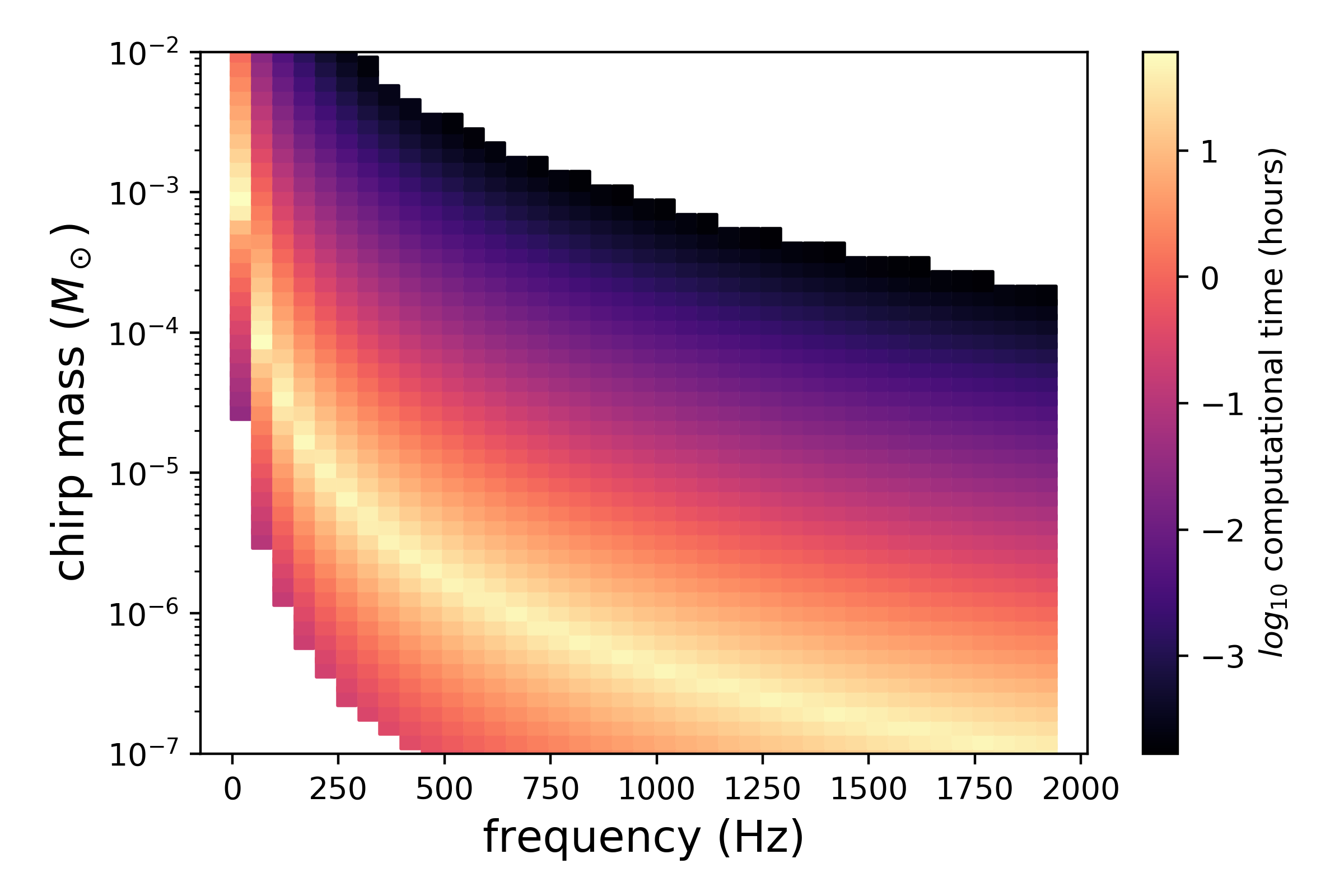}
        % }%
        % \subfigure[]{%
        %   \label{tfftvstobscomp}
        %   \includegraphics[width=0.5\textwidth]{figures/tobs_vs_tfft_comptime_col_dir.jpg}
        % }\\ %  ------- End of the first row ----------------------%
    % \end{center}
    \caption[]{%
We show the computational cost to perform the Generalized Frequency-Hough in a directed search as a function of the source parameter space. We assume a known sky position, so the only limit on the observation time or $\TFFT$ comes from the source's intrinsic properties (its lifetime, its spin-up, etc.). The computational cost varies greatly across this space, with more ``transient'' signals requiring a lot less time than more ``continuous'' ones. White space in the bottom left-hand corner appears because for those parameter points, the linear approximation to equation \eqref{powlaws}, equation \eqref{taylor}, is valid. There are not any points in the right-hand corner because systems with these frequencies and chirp masses would either spin-up too quickly ($\dot{f}>1$ Hz/s) or merge too quickly ($\Tobs<1$ s). We observe a maximum as a ``yellow'' curve because we fix the observation time to be at most one year, and at points below this curve, the Fast Fourier Transform time is increasing, thus resulting in a decrease in computational time. Based on our estimates, on 2000 Quad-Core Intel Core i7 cores in parallel, a directed search using the Generalized Frequency-Hough on one detector's data would take only a few days. 
% Note that creating different peakmaps is not that computationally heavy compared to actually running the Generalized Frequency-Hough. % \cite{longpmr}.
     }%
   \label{dircomp}
\end{figure}

Using the optimal parameters in figure \ref{bestfft}, we estimate the computational cost as a function of the frequency/chirp mass parameter space for a search in 50 Hz bands. For each chirp mass in each frequency band, we evaluate the duration of the signal and the maximum spin-up. We then check to see if the frequency evolution is nonlinear, and exclude signals whose durations are less than 1 second or that have spin-ups greater than 1 Hz/s. The results are shown in figure \ref{dircomp}, where each point corresponds to a Generalized Frequency-Hough transform being performed for a 50 Hz frequency band tuned to that particular chirp mass. To determine the computational time required, we ran the Generalized Frequency-Hough on a peakmap that spanned 50 Hz and 5.5 hours. The primary computational burden of the Generalized Frequency-Hough are two for-loops, one over the number of time steps and one over the number of points in the $k$ grid. In this case, the total number of iterations was around 2 million, and the Generalized Frequency-Hough ran in about 140 seconds on one Quad-Core Intel Core i7, so the time per iteration is around 50 $\mu$s . We then determined the number of total iterations required as a function of the chirp mass and frequency, and multiplied these numbers by the time per iteration. The results are shown in figure \ref{dircomp}. We must also account for the fact that the computation time corresponding to most points in figure \ref{dircomp} is for a specific analysis duration that is less than a year, 
% so the number of times we need to run the Generalized Frequency-Hough is $\Tobs/T_{tot}$ where $T_{tot}=1$ year.
Therefore in total, the computational time required for one Quad-Core Intel Core i7 is around 3400 days. When divided on $\sim 2000$ cores, after peakmaps have been created, a directed search can be performed in only a few days.

We note that the computational cost in figure \ref{compdirparm} is idealistic because it assumes a fixed sky location; in practice, the position of a particular source of gravitational waves is known with a degree of uncertainty, so it is possible that depending on $\TFFT$, that more sky points will be necessary even for a directed search. In this case, the parameter space will have to be limited, but based on the limits forecast in section \ref{pbhlimit}, we do not expect to be able to place constraints on inspirals with very low chirp masses, which also turn out to be the most computationally heavy part of a directed or all-sky search. 

The search design is then clear: we divide the 20-2000 Hz band of our detectors in sub-bands that depend on the chirp mass and frequency evolution of the expected signals in that sub-band, and create time/frequency peakmaps with different $\TFFT$ and $\Tobs$ (these specific values are shown in figure \ref{bestfft}). For each peakmap, we run the Generalized Frequency-Hough, and then follow the post-processing steps in section \ref{pproc}.

If we explore a portion of the parameter space in which the signal is linear, we can use the original Frequency-Hough, and also employ Band Sampled Data files \cite{piccinibsd}, which are more suited to situations in which the signal spans a narrow frequency band. Within this data analysis framework, we can easily change $\TFFT$ to values that exceed the $\TFFT$ chosen in the Short Fast Fourier Transform Databases."

% \begin{eqnarray}
%     dk &\simeq& -nk\frac{\delta f}{f_{\rm max}} \\
%     &\simeq& 6.64\times 10^{-15}\left(\frac{\mathcal{M}}{10^{-3} M_\odot}\right)^{5/3}\left(\frac{32}{\TFFT}\right)\left(\frac{100}{f_{\rm max}}\right)
% \end{eqnarray}

\subsubsection{All-sky search}

For the all-sky search, we must account for the Doppler shift due to the relative motion of the earth and source. In continuous-wave searches, the Doppler shift requires that a shorter $\TFFT$ be taken relative to those presented in figure \ref{bestfft} to confine the frequency modulations to one frequency bin. Furthemore $\TFFT$ is frequency-dependent because the Doppler shift is larger at higher frequencies \cite{Astone:2014esa}. To estimate the computational cost of an all-sky search, we require that $\TFFT<1024$ s, which implies a reasonable number of sky patches to search over \cite{abbott2019searches}. Because much of the parameter space is dominated by transient signals, the spin-up is high and $\TFFT<<1024$ s. While a short $\TFFT$ does not imply a good sensitivity, it does limit the number of sky bins necessary to search over. Sky localization in the first stage of an all-sky search for PBHs is therefore poor compared to standard continuous-wave searches, but the coarseness of our sky-grid makes this search more tractable.

A detailed explanation of the construction of a sky grid for an all-sky search is presented in \cite{Astone:2014esa}. Here we highlight the important parts relevant for our estimation of the computational cost. We first define the maximum frequency shift induced by the Doppler effect as:

\begin{equation}
B=\frac{v_{\rm orb}}{c}f,
\end{equation}
where $v_{\rm orb}$ is the earth's velocity around the sun. $B$ is also known as the Doppler band, with the following number of frequency bins in it $N_D$:

\begin{equation}
    N_D=\frac{B}{\delta f}.
\end{equation}
The number of sky points $N_{\rm sky}$ in the grid related to $N_D$ by:

\begin{equation}
    N_{\rm sky}=4\pi N_D^2,
\end{equation}
where we have set the sky overresolution factor $K_{\rm sky}=1$ (see equation 43 in \cite{Astone:2014esa} for more details on this factor).

\begin{figure*}[ht!]
     \begin{center}
        \subfigure[ ]{%
            \label{histnd}
            \includegraphics[width=0.5\textwidth]{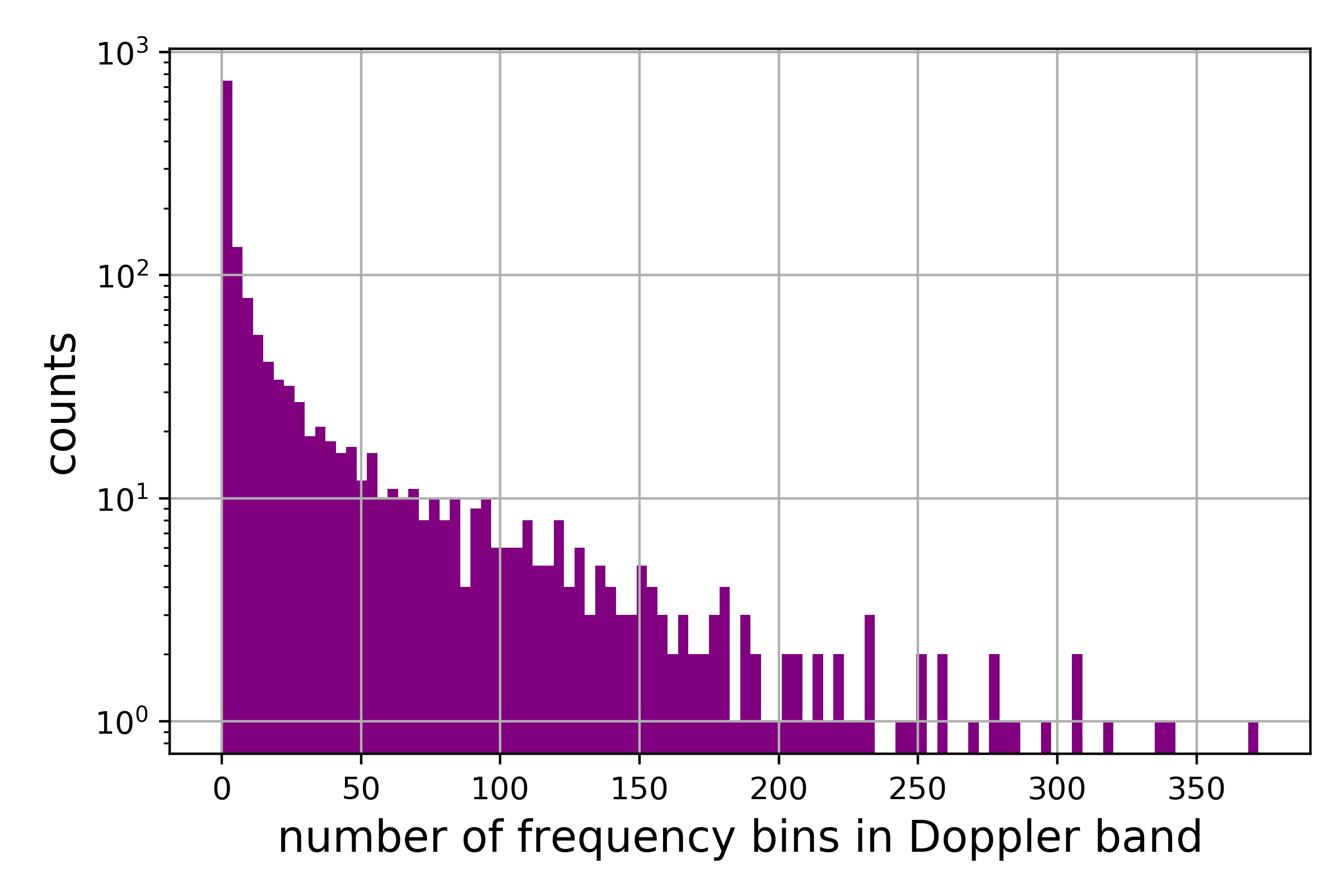}
        }%
        \subfigure[]{%
           \label{skycomp}
           \includegraphics[width=0.5\textwidth]{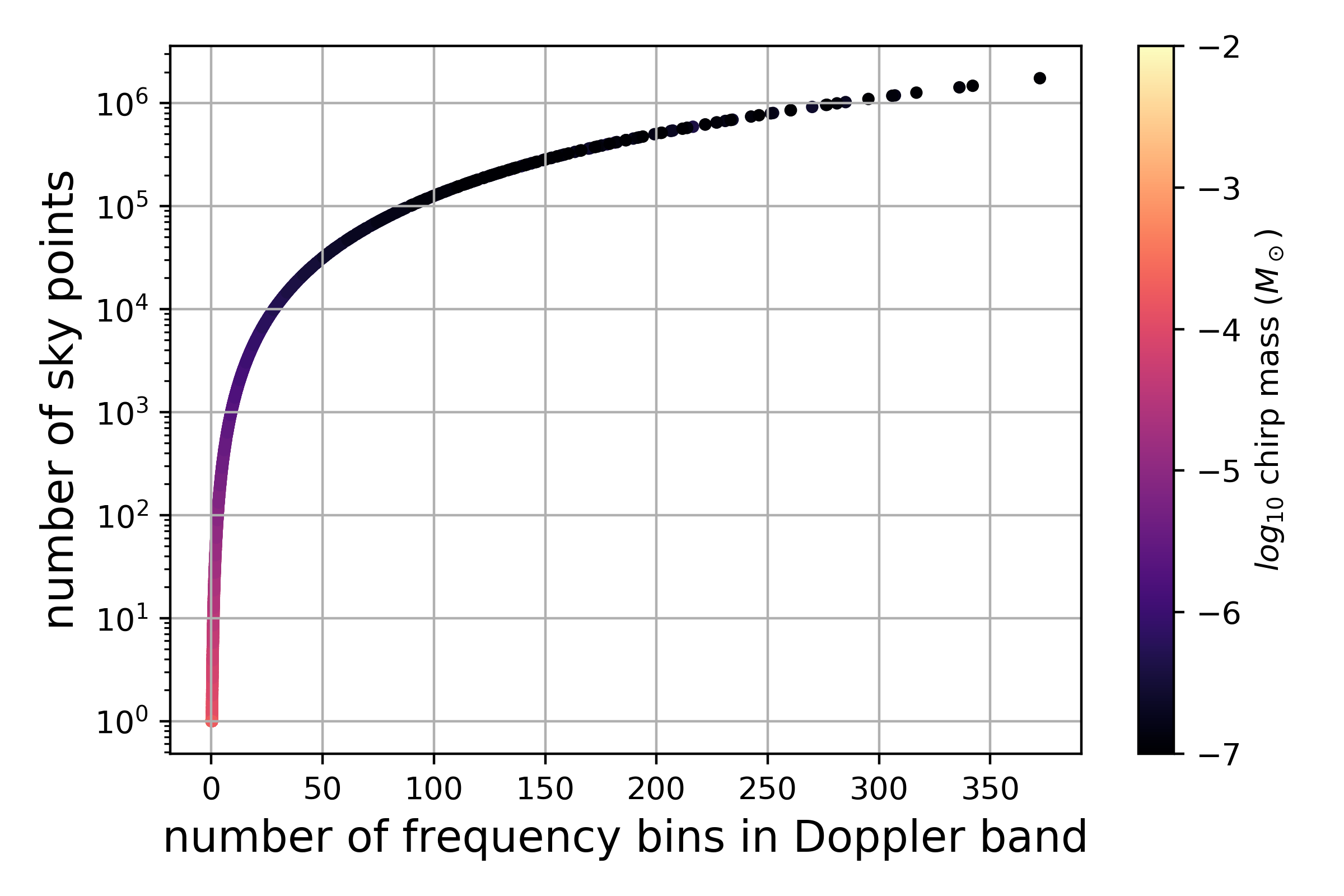}
        }\\ %  ------- End of the first row ----------------------%
%        \subfigure[Transformation of figure \ref{fig:second} to different coordinates]{%
%            \label{fig:third}
%            \includegraphics[width=0.49\textwidth]{figures/time_x_map_sig.jpg}
%        }%
%        \subfigure[Histogram showing injected signal's parameters]{%
%            \label{fig:fourth}
%            \includegraphics[width=0.49\textwidth]{figures/x0_k_hough_map_sig.jpg}
%        }\\ %  ------- End of the second row ----------------------%
%
%        \subfigure[Phase-corrected time/frequency map using \ref{fig:fourth}]{%
%            \label{fig:fifth}
%            \includegraphics[width=0.49\textwidth]{figures/time_freq_map_sig_phase_corr.jpg}
%        }%
%%
%        \subfigure[Hough map, phase-corrected signal's parameters]{%
%            \label{fig:six}
%            \includegraphics[width=0.49\textwidth]{figures/f0_fdot0_hough_map_ph_cor.jpg}
%        }%
%
    \end{center}
    \caption[]{Left: a histogram of $N_D$, the number of frequency bins in each Doppler band, is shown for a hypothetical all-sky search over the parameter space 20-2000 Hz for $\mathcal{M}=[10^{-7},10^{-2}]M_\odot$. Most values of $N_D$ are small because most of the parameter space results in more transient signals, which require shorter $\TFFT$. The lowest chirp masses tend to produce the longest-lived signals, with the lowest spin-ups, which then allow for higher $\TFFT$ and therefore many more frequency bins in the Doppler band. Right: the number of sky points is plotted as a function of $N_D$, with the chirp mass colored. The greatest contributors to the number of sky points in the grid come from the smallest chirp masses at the lowest frequencies in the search parameter space.%.
     }%
   \label{sky_and_hist}
\end{figure*}

For the chirp mass and frequency parameter space, we calculate $N_D$ based on $\TFFT$ in  figure \ref{bestfft}, ensuring $\TFFT<1024$ s, and histogram the values of $N_D$ in figure \ref{histnd}. This histogram shows that the sky grid in most parts of the parameter space does not contain that many points, relative to standard continuous-wave all-sky searches. Indeed, 
% for the parameter space for which the directed search computational cost was estimated (figure \ref{dircomp}), 
we find that a total of $\mathcal{O}(10^7)$ sky points are necessary to be searched over to cover a chirp mass range $[10^{-7}-10^{-2}] M_\odot$ over $20-2000$ Hz. Based on the estimation in section \ref{dirser} (which is essentially an estimate for a fixed sky location), it is computationally infeasible to do a full, comprehensive all-sky search for PBH binaries, unless the parameter space is restricted to higher chirp mass signals.

To understand where this large number of sky points comes from, we show in figure \ref{skycomp} how the sky points are distributed as a function of $N_D$ and $\mathcal{M}$. The greatest contribution to the number of sky points comes from the lowest chirp masses considered at the lowest frequencies. We therefore conclude that an all-sky search covering the full parameter space is not feasible.

% \begin{figure}
%     \centering
%     \includegraphics[width=0.5\textwidth]{figures/hist_ND_pbh.jpg}
%     \caption{A histogram of $N_D$, the number of frequency bins in each Doppler band, is shown for a hypothetical all-sky search over the parameter space 20-2000 Hz and from $\mathcal{M}=[10^{-6},10^{-3}]M_\odot$. Most values for $N_D$ are small, which typical results from the more transient (higher chirp mass) signals. The lowest chirp masses tend to produce the longest lived signals, with the lowest spin-ups, which then allow for higher $\TFFT$ and therefore many more frequency bins in the Doppler band.  }
%     \label{histnd}
% \end{figure}

% \begin{figure}
%     \centering
%     \includegraphics[width=0.5\textwidth]{figures/nsky_vs_Nd_mc_col.jpg}
%     \caption{The number of sky points is plotted as a function of $N_D$, the number of frequency bins per Doppler band, with the chirp mass colored. The greatest contributors to the number of sky points in the grid come from the lowest chirp masses at the lowest frequencies in the search parameter space. }
%     \label{skycomp}
% \end{figure}

\begin{figure*}[ht!]
     \begin{center}
        \subfigure[ ]{%
            \label{theosensh0}
            \includegraphics[width=0.5\textwidth]{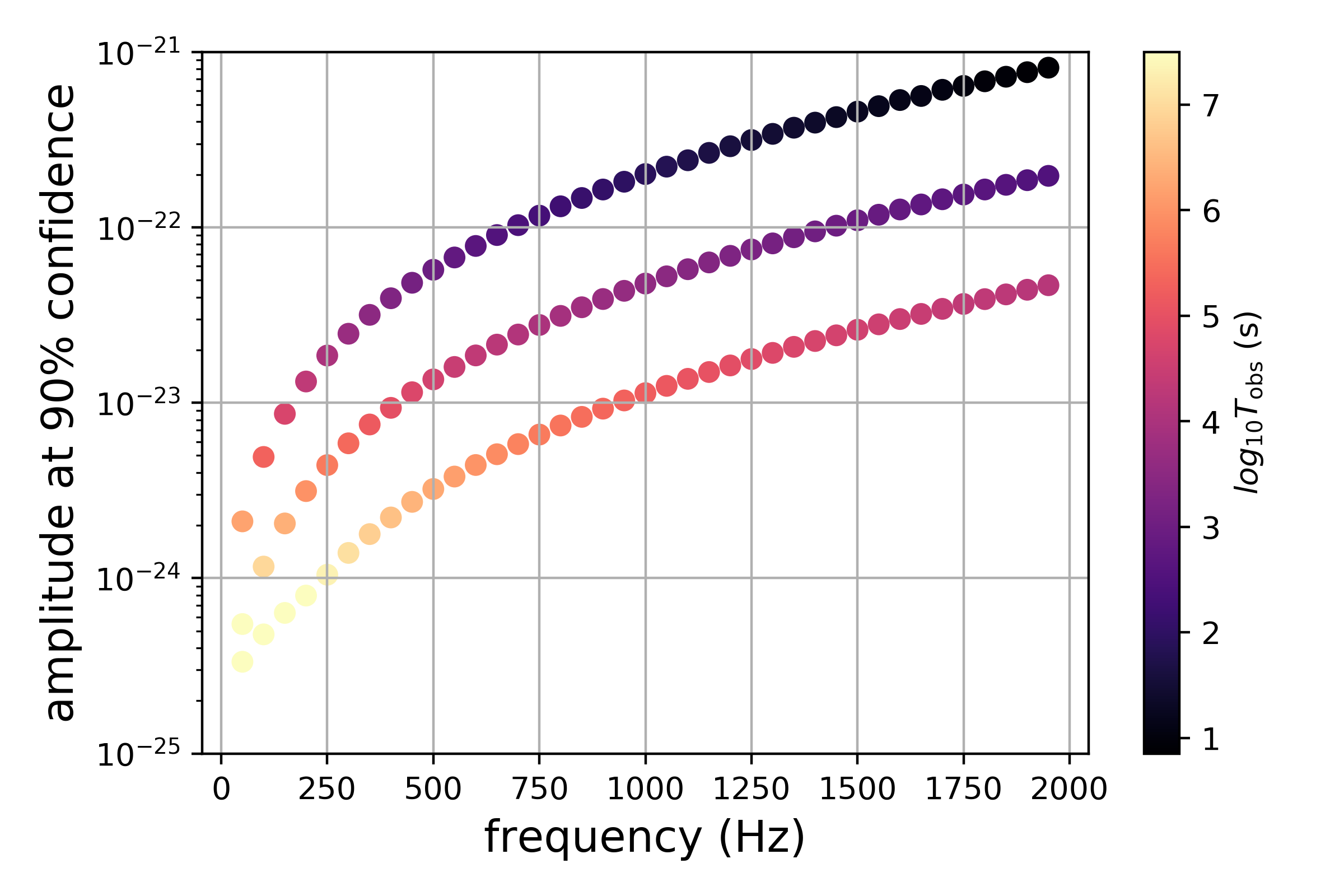}
        }%
        \subfigure[]{%
           \label{theosensd}
           \includegraphics[width=0.5\textwidth]{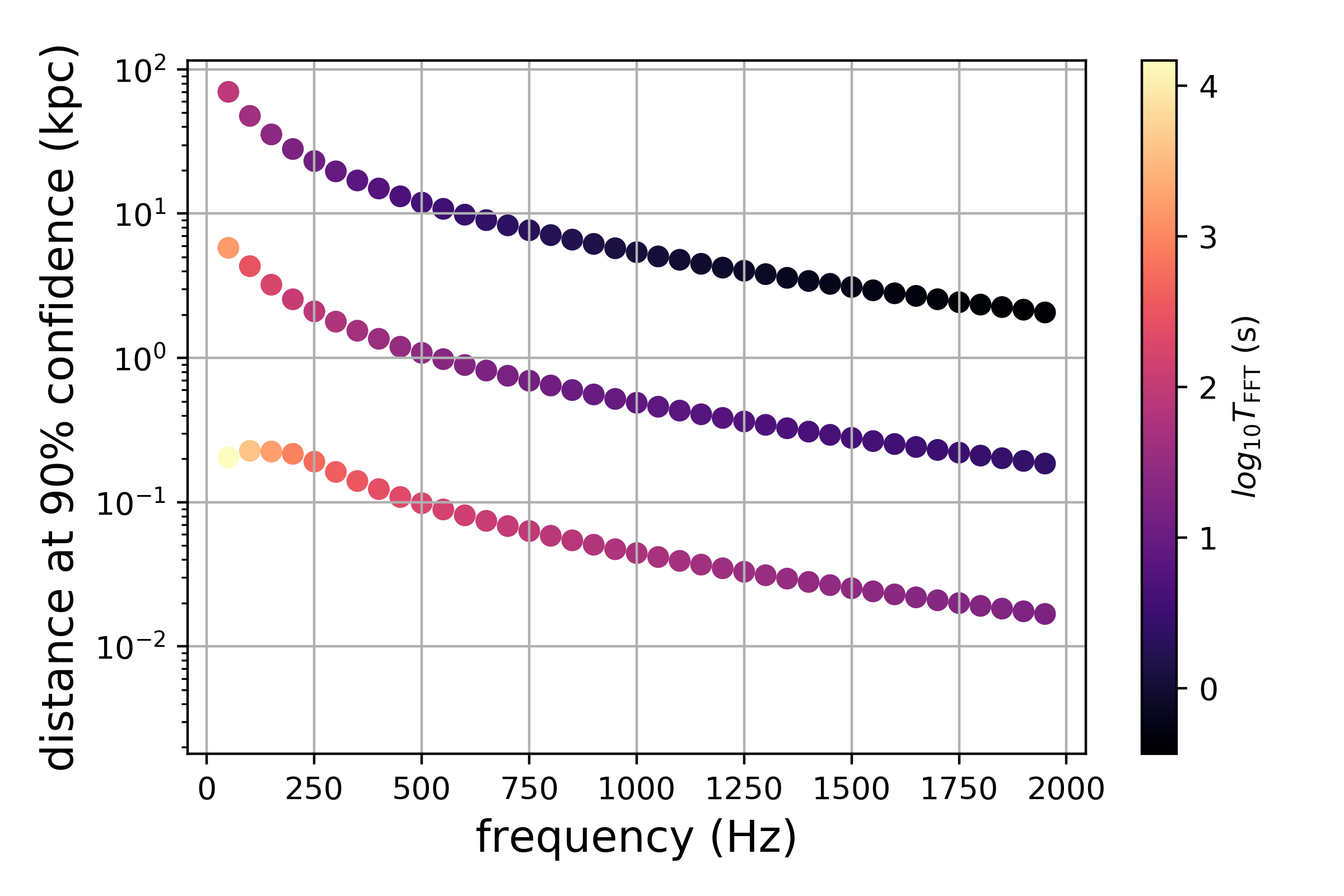}
        }\\ %  ------- End of the first row ----------------------%
%        \subfigure[Transformation of figure \ref{fig:second} to different coordinates]{%
%            \label{fig:third}
%            \includegraphics[width=0.49\textwidth]{figures/time_x_map_sig.jpg}
%        }%
%        \subfigure[Histogram showing injected signal's parameters]{%
%            \label{fig:fourth}
%            \includegraphics[width=0.49\textwidth]{figures/x0_k_hough_map_sig.jpg}
%        }\\ %  ------- End of the second row ----------------------%
%
%        \subfigure[Phase-corrected time/frequency map using \ref{fig:fourth}]{%
%            \label{fig:fifth}
%            \includegraphics[width=0.49\textwidth]{figures/time_freq_map_sig_phase_corr.jpg}
%        }%
%%
%        \subfigure[Hough map, phase-corrected signal's parameters]{%
%            \label{fig:six}
%            \includegraphics[width=0.49\textwidth]{figures/f0_fdot0_hough_map_ph_cor.jpg}
%        }%
%
    \end{center}
    \caption[]{%
   The left-hand plot shows the minimum detectable amplitude $h_0$ at $90\%$ confidence as a function of gravitational-wave frequency in $\fstep$ Hz bands, with the log of the signal duration colored, for three chirp masses: $\mathcal{M}=[10^{-5},10^{-4},10^{-3}]M_\odot$. The right-hand plot shows the distance reach at $90\%$ confidence as a function of gravitational wave frequency, with the corresponding $\TFFT$ colored. We fix the maximum duration of a signal to be equal to the fiducial observation time of one year, though the durations of signals at different frequencies can be less than one year. $\TFFT$ is chosen based on the spin-up of the signal at the end of observation time, hence it varies with frequency. In both plots, the curves from top to bottom correspond to decreasing chirp mass.
     }% [\round \mcthree,\roundtwo \mctwo,\roundtwo \mcone ]
   \label{theo_sens_h0_dist}
\end{figure*}

\subsection{Theoretical sensitivity}\label{theosens}

We determine an analytic expression for the theoretical sensitivity of our method to search for inspiraling PBHs. This section is in fact a generalization of the sensitivity estimate in \cite{PhysRevD.98.102004}.

We assume the signal is periodic but whose frequency and amplitude vary in time. A semi-coherent analysis is based on the condition that in each data segment, of length $\TFFT$, the signal frequency and amplitude are approximately constant. In particular, the frequency does not shift more than one frequency bin.

The same procedure in \cite{PhysRevD.98.102004} can be followed, noting that the amplitude evolution is given by equation \ref{hoft}, which differs from that due to an asymmetrically rotating neutron star. We can then rewrite equations 31 and 32 in \cite{PhysRevD.98.102004} as:

\begin{equation}
h(t_i)=\mathcal{A}f(t_i)^{2/3}=\mathcal{A}\mathcal{F}_i ,
\end{equation}
where $\mathcal{F}_i=f(t_i)^{2/3}$:

\begin{equation}
\mathcal{A}=\frac{4}{d}\left(\frac{G \mathcal{M}}{c^2}\right)^{5/3}\left(\frac{\pi}{c}\right)^{2/3}.
\label{mathA}
\end{equation}
Then the formula for the minimum detectable $\mathcal{A}$ is the same as for transient continuous-wave searches, which we rewrite here for completeness:

% \begin{widetext}
\begin{multline}
    \mathcal{A}_{min}=\frac{4.02}{N^{1/4}\theta_{\rm thr}^{1/2}}\sqrt{\frac{N}{\TFFT}}\left(\sum_i \frac{\mathcal{F}^2_i}{S_n(f_i)}\right)^{-1/2} \\ \times \left(\frac{p_0(1-p_0)}{p^2_1}\right)^{1/4}\sqrt{\left(CR_{\rm thr}-\sqrt{2}\erfc^{-1}(2\Gamma)\right)},
    \label{h00min}
\end{multline}
% \end{widetext}
where  $\theta_{\rm thr}$ is the threshold for peak selection selection in the whitened spectra, $p_0$ is the probability of selecting a peak above the threshold $\theta_{\rm thr}$ if the data contains only noise, $p_1$ = $e^{-\theta_{\rm thr}}-$  2$e^{-2\theta_{\rm thr}}$ $+e^{-3\theta_{\rm thr}}$ , $CR_{\rm thr}$ is the threshold on the critical ratio we use to select candidates in the final Frequency-Hough map, and $\Gamma$ is the chosen confidence level.

The minimum detectable strain at a given confidence level can be obtained from equation \ref{h00min} using a suitable ``frequency'' (indeed $h_{0,min}=\mathcal{A}_{min}\cdot \mathrm{frequency}^{2/3}$). We use the initial frequency $f_0$. 

There are a couple of differences in the interpretation of equation \ref{h00min} when compared to the neutron star case: (1) the $\mathcal{F}_i$ are frequencies that are increasing with time, not decreasing, and (2) $\TFFT$ should now be fixed by the maximum spin-up of the signal which will be at the end time of the analysis (or the time at which equation \ref{chirp_pl} fails to model the inspiral).

The maximum distance reach $d_{\rm max}$ is now obtained
%given 
by combining equations \ref{mathA} and \ref{h00min}:
\begin{widetext}

\begin{equation}
d_{\rm max}=0.995\left(\frac{G \mathcal{M}}{c^2}\right)^{5/3}\left(\frac{\pi}{c}\right)^{2/3} \frac{\TFFT}{\sqrt{\Tobs}}\left(\sum_i \frac{\mathcal{F}^2_i}{S_n(f_i)}\right)^{1/2}\left(\frac{p_0(1-p_0)}{Np^2_1}\right)^{-1/4}\sqrt{\frac{\theta_{\rm thr}}{\left(CR_{\rm thr}-\sqrt{2}\erfc^{-1}(2\Gamma)\right)}}.
\label{dmax}
\end{equation}
\end{widetext}
The braking index $n=11/3$ enters into equation \ref{dmax} through the frequency dependence in $\mathcal{F}_i$. For different choices of observation time and chirp mass, we will have different maximum distance reaches.

Figure \ref{theosensh0} shows how observation time affects the strain amplitude induced on the detector as a function of frequency. For lower chirp masses, $h_0$ is lower, but we can integrate for longer times because the signal is more ``continuous'' than those for higher chirp masses. For higher chirp masses, the signals are transient, lasting for much shorter amounts of time, but are intrinsically stronger, though they induce smaller strains on the detector because they do not last as long. The scenario that is truly better must be weighted by the probability of PBH mergers as a function of chirp mass and the kinds of constraints we can place on the PBH mass fraction, which will be discussed in section \ref{pbhlimit}.

Figure \ref{theosensd} shows the maximum distance reach for different chirp masses as a function of frequency. Additionally, $\TFFT$ is colored, which shows how the spin-up of the inspiraling system affects the sensitivity of the search. For longer signals, the distance reach is smaller, but the strain sensitivity is actually better, because we can use longer $\TFFT$ and observe for longer times.
% But some chirp masses in this regime can be probed with the Frequency-Hough or another continuous-wave technique. 
However, PBH systems with greater chirp mass can actually be seen farther away from us, even though they induce a smaller strain on the detector, because the theoretical distance reach is proportional to $\mathcal{M}^{5/3}$. 
% Moreover, the number of expected PBH mergers is higher for lower chirp masses around $10^{-5}M_\odot$ than for higher ones, around $10^{-3}M_\odot$.

\section{Limits on PBH abundance} \label{pbhlimit}

\begin{figure}[t]
    \centering
    \hspace{-3mm} 
    \includegraphics[width=\columnwidth]{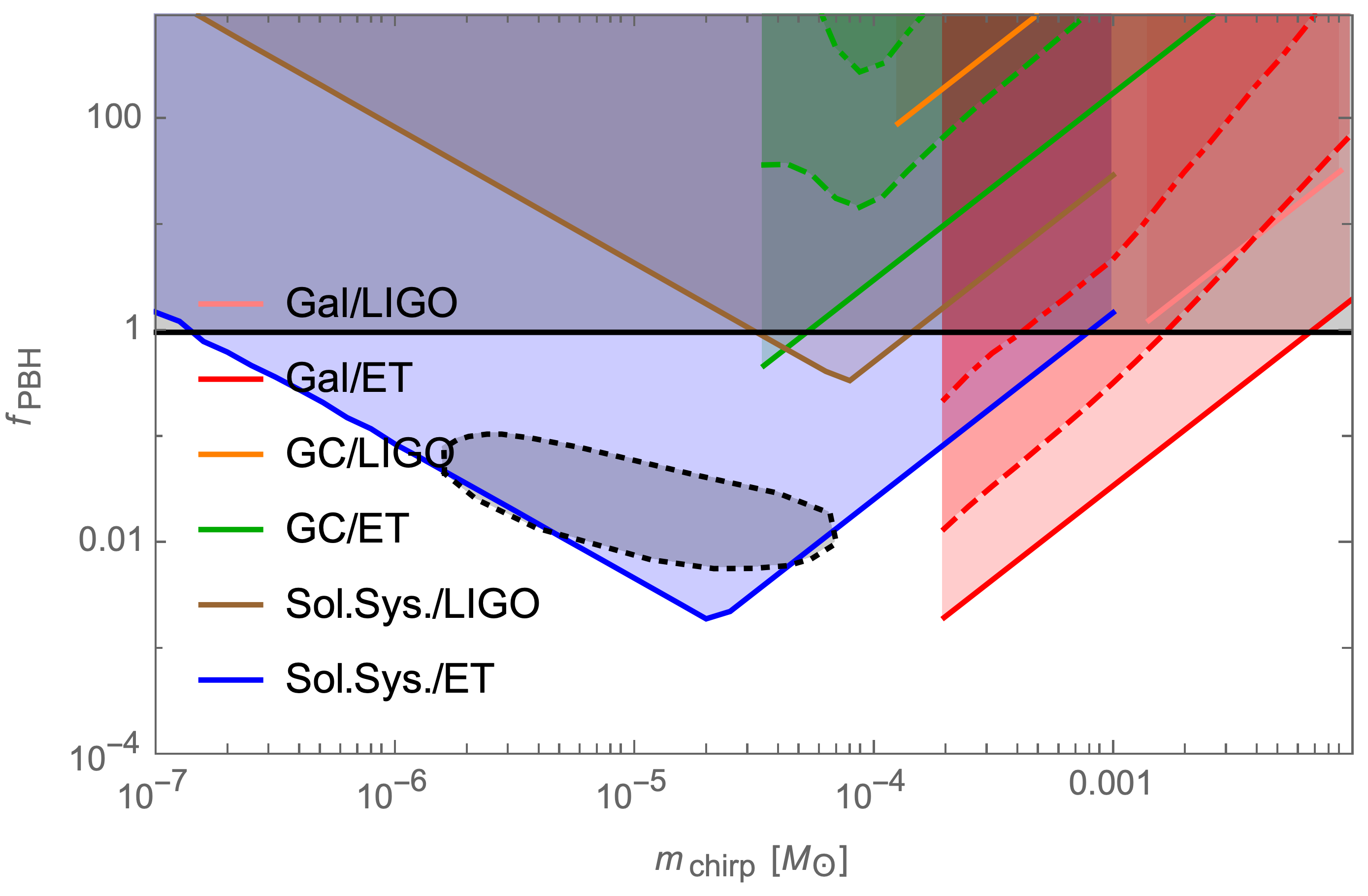}
    \caption{Expected limits on the dark matter fraction made of PBHs as a function of the chirp mass, for primordial binaries in Case 1 - agnostic mass function - (solid lines) and Case 2 - thermal mass function - (dashed lines), and for binaries formed through tidal capture in Case 2 (dotted-dashed lines).  The different colors represent the limits from galactic binaries (gal), from the galactic center (GC) and in the solar system vicinity (sol. sys.), for the expected sensitivities of advanced LIGO/Virgo and ET. The dotted elliptic region represents constraints from the Optical Gravitational Lensing Experiment (OGLE) and the Subaru Hyper Suprime-Cam (HSC) \cite{Niikura:2019kqi} for comparison. }\label{fig:limits}
\end{figure}

We forecast limits on the fraction of dark matter that could be composed of PBHs. Based on figure \ref{theosensd} and equation \ref{dmax}, we calculate the expected constraint on $f_{\rm PBH}$ such that at a particular distance and frequency, we would obtain one event for a specific merging rate scenario in a given observation time. We show such limits in Figure~\ref{fig:limits}, and find that LIGO/Virgo can already constrain $f_{\rm PBH} \lesssim 1$ at chirp masses between $4 \times 10^{-5} M_\odot$ and $10^{-3} M_\odot$, within solar system vicinity (up to 2 kpc distance), or from the galactic center. Farther away binaries in the galactic halo may also be probed with the method.  We also find that Einstein Telescope will be able to set limits between $10^{-6}$ and $10^{-2} M_\odot$, with optimal limits $f_{\rm PBH} \lesssim 10^{-2}$ between $10^{-4}$ and $10^{-3} M_\odot$. 
 The most interesting limits are obtained for primordial binaries, whose merger rates are however controversial.  For binaries formed by tidal capture, a relevant limit can only obtained for galactic binaries with ET sensitivity.  Our constraints are complementary to those in \cite{Wang:2019kaf}, in which the imprint in the stochastic gravitational-wave background due to PBH mergers is calculated for PBH masses of $[10^{-8},1]M_\odot$.

{Continuous waves are therefore promising probes of sub-solar PBHs that will complement star, quasar and supernovae microlensing searches.  In particular, ET will probe the interesting region between $10^{-6}$ and $10^{-5} M_\odot$, in which a series of microlensing events have been discovered in OGLE observations towards the galactic center~\cite{Niikura:2019kqi}, suggesting $f_{\rm PBH} \sim 10^{-2}$ in this range.  Outside of this range, other microlensing limits exist, but they rely on the assumption that PBHs are uniformly distributed in the galactic halo. These limits can also be evaded if instead PBHs are clustered~\cite{Carr:2019kxo}.  This case is the most relevant one for gravitational-wave observations, and clustering plays a crucial role in the determination of PBH merging rates.  Continuous waves could therefore help to disentangle the different effects linked to PBH clustering.  While microlensing surveys towards the Magellanic clouds or M31 set limits on the uniform fraction of PBHs, continuous waves will probe PBHs in clusters, which make the two methods complementary to each other. }

\section{Conclusions} \label{concl}

In this paper we presented the first adaptation of continuous-wave methods to directly detect inspiraling planetary-mass binaries, and performed new calculations for the rates of these inspirals for different PBH mass functions. Our work shows that 
if such PBHs make up most of the dark matter,
%even if PBHs make up only a small fraction of dark matter, 
it is possible to probe a wide parameter space of PBH chirp masses with continuous-wave techniques, between $10^{-6}$ and $10^{-3} M_\odot$. In current and future observation runs of LIGO/Virgo, we will be able to place constraints, or even make a detection, of PBH inspirals within our galaxy, if they exist. We have also forecast some expected constraints on the fraction that PBHs compose of dark matter, assuming different mass functions, formation mechanisms and distances from us.

We have shown that traditional continuous-wave methods alone are not suited to probe an important fraction
%cannot probe a good portion 
of the possible PBH inspiral parameter space. Transient continuous-wave techniques, developed originally to search for post-merger remnants of supernova or binary neutron star mergers, are in fact quite effective to search for PBHs at galactic distances. The combination of both methods, however, is ideal: continuous-wave techniques can see very near to us, at very low PBH chirp masses, while transient continuous-wave methods can see farther out, for higher chirp masses.

In the future we plan to run a real search on data from LIGO/Virgo's third observation run using the method and search design presented here, pointing towards the galactic center.

We have fully characterized this method: we have theoretically estimated its sensitivity, determined an ideal way to run a search by weighing gains in observation time against losses in sensitivity due to a rapidly spinning up signal, assessed the computational cost of both a directed and an all-sky search, and characterized the accessible parameter space in terms of our analysis parameters.

It is worth noting that the adaption of continuous-wave techniques is not just limited to the current detection era. Indeed, our work will have implications for PBH inspiral detection in ET that will probe
%: we will be able to probe even 
smaller chirp masses (lower frequencies) and lower abundances than 
%we can with 
LIGO/Virgo currently can. Moreover, LISA will come online within 15 years, and since many sources detectable by LISA are expected to be long-lived and have extremely small spin-ups, our methods can also be applied to find continuous waves from binary white dwarfs, inspiraling supermassive black holes, etc. Characterizing the LISA sources to which our method can be sensitive is the subject of future work.

\section*{Acknowledgements}

We would like to thank the continuous wave and dark matter groups within the LIGO/Virgo collaboration, as well as Bernard Whiting and Cristiano Palomba, for very useful discussions.

We would also like to thank the Rome Virgo group for the tools necessary to perform these studies: the Short Fast Fourier Transform Databases and the Band Sample Data structures, and for developing the methods used here. Additionally we would like to thank Luca Rei for managing data transfer between Rome, CNAF and Louvain.

Computational resources have been provided by the supercomputing facilities of the Université catholique de Louvain (CISM/UCL) and the Consortium des Équipements de Calcul Intensif en Fédération Wallonie Bruxelles (CÉCI) funded by the Fond de la Recherche Scientifique de Belgique (F.R.S.-FNRS) under convention 2.5020.11 and by the Walloon Region.

This research has made use of data, software and/or web tools obtained from the Gravitational Wave Open Science Center (https://www.gw-openscience.org/ ), a service of LIGO Laboratory, the LIGO Scientific Collaboration and the Virgo Collaboration. LIGO Laboratory and Advanced LIGO are funded by the United States National Science Foundation (NSF) as well as the Science and Technology Facilities Council (STFC) of the United Kingdom, the Max-Planck-Society (MPS), and the State of Niedersachsen/Germany for support of the construction of Advanced LIGO and construction and operation of the GEO600 detector. Additional support for Advanced LIGO was provided by the Australian Research Council. Virgo is funded, through the European Gravitational Observatory (EGO), by the French Centre National de Recherche Scientifique (CNRS), the Italian Istituto Nazionale della Fisica Nucleare (INFN) and the Dutch Nikhef, with contributions by institutions from Belgium, Germany, Greece, Hungary, Ireland, Japan, Monaco, Poland, Portugal, Spain.

We also wish to acknowledge the support of the INFN-CNAF computing center for its help with the storage and transfer of the data used in this paper.

We thank the anonymous referees for their comments, which have improved the paper.

We would like to thank all of the essential workers who put their health at risk during the COVID-19 pandemic, without whom we would not have been able to complete this work.

A.L.M. is a beneficiary of a FSR Incoming Post-doctoral Fellowship. F.D.L. is supported by a FRIA grant from the Fonds de la Recherche Scientifique FNRS, Belgium.

\bibliographystyle{ieeetr}
\bibliography{biblio}

\end{document}